\documentstyle[preprint,prb,aps,epsfig]{revtex} 
\begin{document}
\tighten
\title{An Electromechanical Which--Path Interferometer}
\author{A. D. Armour\cite{auth}} 
\address{The Blackett Laboratory, Imperial College of Science,
Technology and Medicine, London SW7~2BZ, United Kingdom}
\author{M. P. Blencowe\cite{auth2}}
\address{Department of Physics and Astronomy, Dartmouth College,
Hanover, New Hampshire 03755}
\date{\today}
\maketitle
\begin{abstract}
We investigate the possibility of an electromechanical which--path 
interferometer, in 
which electrons travelling through an 
Aharonov--Bohm  
ring incorporating a quantum dot in one of the arms are dephased by an interaction 
with the fundamental flexural mode of a 
radio frequency cantilever. The cantilever is positioned so that its tip
 lies just above the dot and a bias is applied so that an electric field
  exists between the dot and the tip. This electric field is modified when an 
  additional electron hops onto the dot, coupling the 
   flexural mode of the cantilever and the microscopic electronic 
  degrees of freedom. We analyze the transmission properties of this system 
  and the dependence of interference fringe visibility on the cantilever--dot 
  coupling and on the mechanical properties of the cantilever. 
  The fringes are progressively 
  destroyed as the interaction with the cantilever is turned up, in 
  part due to dephasing arising  from the entanglement of the electron and 
  cantilever states and also due to the thermal smearing that results from 
  fluctuations in the state of the cantilever. When the dwell time of the 
  electron on the dot is comparable to or longer than the cantilever period,
  we find coherent 
  features in the transmission amplitude. These features are washed out 
  when the cantilever is decohered by its coupling to the environment. 
  \end{abstract}

\section{Introduction}

Which--path devices such as the canonical
 two--slit interference experiment, where a  measurement is made of the
path a particle takes,  
 have a long 
history going back as far as the early debates about 
complementarity.\cite{hist,Fey} 
Recent interest in their investigation has been stimulated
partly by advances in experimental
techniques, which have lead to the realization of several different
varieties of which--path systems in the laboratory,\cite{Buks,op} and
partly by accompanying developments in the theory of quantum
measurement.\cite{SS,Scully,Walls} However, it is  the
realization that a which--path experiment provides a very convenient
model system for developing and testing fundamental ideas about
decoherence in mesoscopic systems,\cite{Stern,hacr} that has  increased the
level of interest within the solid state physics community in particular.

Which--path experiments in solid state systems were recently pioneered
by Buks {\emph{et al.}}\cite{Buks} The solid state analog
 of the two--slit interference experiment is the measurement of the oscillations
  in the current passing  through an Aharonov--Bohm (AB) ring as a function 
  of the applied magnetic field. The path taken by an electron may be probed by 
  placing a measuring device close to one of the arms. Buks {\emph{et al.}}
   incorporated a quantum dot in one of the arms in order to slow the electrons 
   down, with a neighboring quantum point contact (QPC) serving as a 
   which--path detector.
An electron travelling around the arm of the ring containing the dot 
 dwells on the dot for a finite amount of time before 
moving on. The presence of the dot alone does not destroy 
the coherence of the electron transport through the ring, so long as the dwell time 
of the electrons is sufficiently short,\cite{Yac} but it does provide time for the 
electron to interact with an external measuring device. A QPC adjacent to the dot 
functions as a measuring device since it can be biased so that its 
conductance is very sensitive to changes in the occupancy of the dot: the 
passage of an electron via the path including the dot leaves behind which--path 
information in the QPC device, although actual knowledge of which path 
an electron took, so--called `true which--path information', can only be obtained 
via further measurement.\cite{Hack} In accord with theoretical
predictions,\cite{Le,Al,Gurvitz} the experiment demonstrated that
 the interference fringes are degraded when the interaction 
between the electrons and the measuring device is sufficiently 
sensitive
for information to be obtained that would help, even in principle,
 to determine which of the two possible paths an electron
took.    

The experiment of Buks {\emph{et al.}} has close parallels with 
the well--known thought experiment (see, e.g., Sec. 1-6 of Ref.\ 
\onlinecite{Fey}), in which 
a light source
is used to detect through which slit an electron passes in a  two--slit
interference experiment. In both cases dephasing is effected by
scatterers, each of which interact once, and once only, with the interfering
particle and whose interactions with each other may safely be ignored.
For the electron--light scattering scheme,\cite{Fey} the scatterers are photons
which probe the electron's state directly, whereas in the
solid--state experiment the scatterers are electrons in the QPC
which interact with the electron on the dot via electrostatic coupling. 
However, because the electron has a finite dwell time on the
dot, it has time to interact with  more than one scatterer in the QPC, 
so that in this case dephasing can be achieved via a series of 
very weak interactions rather than a single strong event, as is the
case in the photon thought experiment.

The present work is concerned with a variation on the system 
considered by Buks 
{\emph{et al.}}, 
in which a radio frequency mechanical cantilever is used, rather than a
QPC, to determine which path an electron takes. A coupling  between electrons 
residing on the dot (which is again on one of the AB interferometer arms) and 
 the
cantilever, whose tip is suspended over the dot, can be set up by developing a 
uniform electric field between the tip of the cantilever and the base of the dot. 
Electrons on the dot couple approximately linearly to the cantilever position, thus leading 
to a coupling between the flexural phonon modes of the cantilever and the
occupancy of the dot. Furthermore, because the coupling strength decays rapidly 
with increasing frequency  
and  also the flexural mode spectrum of a cantilever 
is quite sparse, it turns out that for micron scale  cantilevers  only the fundamental
flexural mode is relevant. Therefore, at low 
temperatures the cantilever can be
treated as a single quantum mechanical oscillator.

As a consequence of the cantilever having effectively only 
one degree of freedom,  the electromechanical which--path 
interferometer exhibits qualitatively different behavior from the QPC 
which--path device. In particular, we 
find that the dephasing behavior of 
the electron due to the cantilever depends on the 
relative magnitudes of the electron dwell time on the dot and the cantilever 
period. When the  dwell time is short compared to the cantilever 
period, the dephasing occurs in a way analogous to Einstein's recoiling slit
thought--experiment.\cite{hist,SS} In contrast, when the dwell time of
the electron on the dot is comparable to or longer than the period of the 
cantilever,
a description of the dephasing in terms of the entanglement of the
cantilever and electron states becomes more appropriate than a
semiclassical picture of momentum transfer.
The effectively harmonic nature of  the cantilever motion
means that the degree of entanglement between the cantilever and 
electronic states
must be periodic, so long as the cantilever interacts only with 
the electron on the dot. If the dwell time
 of the electrons on the dot could be tuned to the cantilever period, 
 then we would 
   be able to erase all which--path information held in the state of the 
   cantilever. 
   This would give us direct 
    evidence that the 
   coupled cantilever and dot were behaving as a single coherent quantum system. 
 However, in practice electron dwell times on a quantum dot have a distribution
  of values and so we can only obtain indirect evidence for the quantum 
 coherence of the  cantilever--dot system, such as the coherent exchange of 
  energy quanta between the electron and the cantilever which give rise to 
 side resonances in the elastic transmission amplitude of the device. 
 
 The environment of the cantilever influences its behavior  in two important ways: 
 over short timescales it destroys the cantilever's phase coherence, whilst
 over longer timescales it damps the cantilever's motion.
 The electromechanical which--path system we propose acts as a probe of 
 the cantilever's decoherence due to it's interaction
  with the environment. We show that the cantilever's decoherence 
  inhibits the coherent
  exchange of energy between the cantilever and the dot and hence manifests 
  itself by washing out the side resonances in the elastic transmission 
  amplitude.

The outline of this paper is as follows. In Sec. II we give our basic 
model for the electromechanical
which--path device and show that the interference properties of the AB
ring, modified to include the dot and cantilever, can be obtained by calculating
the elastic transmission amplitude of the arm containing the dot. In 
Sec. III 
we describe a simple tight--binding model for an electron on a  dot which is linearly
coupled to the cantilever (treated as a quantum oscillator) and to
 propagating states in the leads. We show that recent work on
inelastic resonant tunneling can be adapted to obtain the elastic
transmission amplitude. 

We present results for our model in Sec. IV, where we examine
dephasing as a function of the dot--cantilever coupling in both
the regime where the electron dwell time is short compared to the
cantilever period and  where they are comparable. 

In Sec. V we examine the influence of the cantilever's
environment on the  dephasing produced by the cantilever when the
dwell time is comparable to the period of the cantilever. We obtain a modified
expression for the transmission through the arm with the dot when the 
cantilever is coupled to an environment which consists of a bath of oscillators.

In Sec. VI we outline the principal practical constraints on an
electromechanical which--path device and discuss its feasibility 
using currently available technology. We draw our conclusions in Sec. 
VII and discuss ways in which the present work could be extended.
Appendix A contains a detailed, classical analysis of the cantilever--dot 
coupling and the flexural mode spectrum of the cantilever which underpin 
the simplified model employed in the main text. Finally, appendix B contains 
details of a general calculation of the effect of the cantilever environment on
the transmission amplitude of the dot. 

\section{Theoretical Model}

The ultimate goal of our analysis of the electromechanical which--path device
is to obtain an expression for how
the magnetoresistance of the AB ring varies as a function of the coupling
between the cantilever and an electron on the dot. In particular, we
want to determine how  the amplitude and phase of the
oscillation in the current through the device as the magnetic field is
varied (i.e., the interference fringes) depend on the voltage between the
cantilever and the dot. 
In order to simplify the analysis, we will consider only the case
where the magnetoresistance is measured as an average over a 
time  much longer than any other timescale in the
problem.    
   
Our model builds  on the theoretical analysis of the
QPC which--path experiment,  carried out by a number of
groups.\cite{Le,Al,Gurvitz}  However,
there are also close parallels between the system we
consider and a quantum optical system analyzed recently,\cite{BJK,Bose} in which 
radiation is used to drive an oscillator--mounted mirror into non--classical states. 

Since we are interested in the behavior of the system averaged over
time, it is sufficient to work within the Landauer framework, so that
our task of obtaining the average magnetoresistance is equivalent to
calculating the transmission characteristics of the device.\cite{Datta}
The simplest way of thinking about the transmission probability through  
 the AB ring is in terms of a two--slit experiment: the total 
transmission probability, $\mathcal{T}$,  is given by the coherent sum of 
amplitudes for 
transmission through the left and right arms, $t_l$ and $t_r$,
\begin{equation}
{\mathcal{T}}=|t_l+t_r|^2.\label{sm}
\end{equation}
This approach was employed by Yacoby {\emph{et al.}}\cite{Yac} in their 
analysis of the first experiments on AB rings with a quantum dot 
included in one arm. However, in that case this simple 
formalism proved
inadequate as it neglects the contribution of electrons which have
undergone multiple reflections. The contribution to the transmission
of multiply reflected electrons leads to the enforcement of the
Onsager--type relation for the conductance from source (S) to 
drain (D): $G_{DS}(B)=G_{DS}(-B)$, which must apply for a
two--terminal device.\cite{But1,But2,Imry,Yac2}
If the AB ring is modified to absorb electrons which are scattered 
backwards in the device, then the Onsager--type condition is no longer satisfied 
and the simple--minded formula (\ref{sm}) is valid. Buks 
{\emph{et al.}} incorporated  just such a modification in 
their AB ring and we shall limit our analysis to this case.    
 We write 
$\tilde{t}_{\mathrm QD}$ for the transmission through the arm containing the dot and 
$t_0$ for the other amplitude. The presence of a magnetic flux, $\Phi$, induces 
an additional  relative phase shift between the two paths and so the 
transmission probability in the absence of the cantilever can be written as
\begin{equation}
{\mathcal{T}}=\left|t_0+\tilde{t}_{\mathrm QD}{\mathrm{e}}^{i 
2\pi\Phi/\Phi_0}\right|^2
={\mathcal{T}}^{(0)}+2{\mathrm{Re}}\left[ t^*_0\tilde{t}_{\mathrm QD}{\mathrm{e}}^{i 
2\pi\Phi/\Phi_0}\right], \label{T}
\end{equation}
where $\Phi_0$ is the flux quantum and ${\mathcal{T}}^{(0)}$ is the 
flux independent term of the transmission probability. In practice, 
the transmission through the arm with the dot is much
less than that through the arm without the dot,  
$|\tilde{t}_{\mathrm QD}|\ll|t_0|$, and so ${\mathcal{T}}_0\approx |t_0|^2$.

When there is a non--zero interaction between the cantilever and the
electron on the dot, we must  explicitly include the fact that the
transmission depends on the initial  state of the
cantilever. Since the cantilever contains a macroscopic number 
of atoms we assume that
its initial state can always be described as a thermal mixture. Such a
procedure will necessarily lead us to neglect short time correlations
between the cantilever states, but this will be unimportant if the
magnetoresistance is averaged over a time which is long compared with
the characteristic timescale for the thermalization of the
cantilever's motion.\cite{steady} With the initial state of the 
cantilever assumed to be thermal we have,
\begin{equation}
{\mathcal{T}}=\sum_i\sum_f \rho_i\left[ \langle i|t_0 
+\hat{t}_{\mathrm QD}{\mathrm{e}}^{i 2\pi\Phi/\Phi_0} | f\rangle \left(\langle
  i|t_0 +\hat{t}_{\mathrm QD}{\mathrm{e}}^{i
2\pi\Phi/\Phi_0} | f\rangle\right)^*\right], \label{T2} 
\end{equation}
where $\hat{t}_{\mathrm QD}$ is an operator on cantilever states only, $\rho_i$ is
the usual thermal weight ($\rho_i={\mathrm{e}}^{-\beta
  \epsilon_i}/\sum_j{\mathrm{e}}^{-\beta \epsilon_j}$, with
$\beta=1/k_{\mathrm{B}}T$ and $\epsilon_i$ the energy of the state),
 and we have made no  assumption about the final state
  of the cantilever mode. Because only the dot arm interacts with the cantilever
the interference term is diagonal in the cantilever modes,
\begin{equation}
2{\mathrm{Re}}\left[ \sum_i \rho_i t^*_0\langle i|\hat{t}_{\mathrm QD}|i 
\rangle {\mathrm{e}}^{i 2\pi\Phi/\Phi_0}\right]=
2{\mathrm{Re}}\left[t^*_0\langle\tilde{t}_{\mathrm QD}\rangle 
 {\mathrm{e}}^{i 2\pi\Phi/\Phi_0}\right],
\end{equation}  
where $\langle \tilde{t}_{\mathrm QD}\rangle$ is averaged over the cantilever thermal
state. Thus, only elastic scattering processes contribute to the 
interference term. 

At finite temperatures we also have to average the transmission amplitude for 
transport through the arm containing the dot over the Fermi 
distribution:\cite{Datta}
\begin{equation}
\langle\tilde{t}_{\mathrm QD}\rangle
=\int_{0}^{\infty} {d}\epsilon\left( -\frac{\partial f}{\partial 
\epsilon}\right) \langle t_{\mathrm QD}(\epsilon)\rangle. \label{therm}
\end{equation}

The knowledge of the transmission amplitude allows us to obtain the
amplitude of the periodic oscillations in the current, or in the
language of a generic two--slit experiment, the visibility of the
interference fringes. The visibility, $v$, in a two--slit experiment is 
 defined in terms of the maximum and minimum signal (in this case 
current), measured at the peaks and valleys of the fringes,
 respectively:\cite{vis}
\begin{equation}
v=\frac{{\mathrm{max.}}-{\mathrm{min.}}}{{\mathrm{max.}}+{\mathrm{min.}}}.
\end{equation}

In the case of the electromechanical which--path device considered here, the 
current is proportional to the transmission probability, $\mathcal{T}$, 
given in Eq. (\ref{T2}) above. We can write the transmission amplitudes through 
the two arms in modulus--argument form, 
$\langle\tilde{t}_{\mathrm QD}\rangle=|\langle\tilde{t}_{\mathrm QD}
\rangle|{\mathrm{e}}^{i\alpha}$ and 
${t}_0^*=|t_{0}|{\mathrm{e}}^{-i\beta}$, so that the interference 
part of the transmission probability takes the form
\begin{equation}
2{\mathrm{Re}}\left[t_0^*
\langle\tilde{t}_{\mathrm QD}\rangle{\mathrm{e}}^{i2\pi\Phi/\Phi_0}\right]
=2|{t}_0||\langle\tilde{t}_{\mathrm QD}\rangle|\cos\left( 2\pi\Phi/\Phi_0+ \alpha-\beta 
\right).
\end{equation}
Hence in this case, the visibility of the fringes is
\begin{equation}
v=\frac{2|{t}_0||\langle\tilde{t}_{\mathrm QD}\rangle|}{{\mathcal{T}}^{(0)}}
\simeq \frac{2|\langle\tilde{t}_{\mathrm QD}\rangle|}{|{t}_0|}. 
\end{equation}
The phase of the
transmission amplitude  determines the phase of the fringe
pattern as a function of the magnetic field. Any change in the phase
of the transmission amplitude should therefore be detectable as a change in the
phase of the whole interference pattern. 

In the absence of the dot, and for the ideal case of a ring in which both arms 
are identical, the transmission amplitudes for both arms are 
the same so that ${\mathcal{T}}_0=2 |{t}_0|^2$ and the visibility is unity. 
This demonstrates the utility of $v$ as a measure:  it does not depend 
on the value of the total transmission probability through the device,
but just on its interferometric properties.

In practice, a fringe visibility close to unity cannot be
achieved. Apart from the obvious difficulty in constructing a ring in which both 
arms are identical, there are two further reasons why the `intrinsic' 
visibility is less than unity.\cite{Yac} Firstly, there is more than one conduction 
channel in each arm and so the transport is never really a `one electron' problem. 
Secondly, thermal smearing can play an important r\^{o}le, since the thermal 
smearing length is typically comparable to the size of the ring at temperatures
 of around 100mK.\cite{Yac} However, for our device 
 the reduction in fringe visibility arising from sources 
other than the cantilever is irrelevant: all we 
require is that there should be a measurable change in the visibility of the 
fringes as the electric field coupling the dot and cantilever is turned on.

\section{Transmission Amplitude}

In order to obtain the transmission amplitude for the arm of the AB
 ring containing the dot, we employ a standard tight--binding model for the dot 
and the leads to which it is coupled. In
the Coulomb blockade regime, the quantum dot is modelled by a single localised 
state, at energy $\epsilon_0$, which is coupled to sets of non--interacting,
 propagating, states in
 the leads. The cantilever is treated as a single quantum oscillator mode of frequency
  $\omega_0$, although the generalisation to include a spectrum of modes is 
  straightforward, as we discuss below.   
The total Hamiltonian of the interferometer arm can be written as the
sum of two parts 
\begin{equation}
{\mathcal{H}}={\mathcal{H}}_0+{\mathcal{H}}_1 \label{Ham1},
\end{equation}
with non--interacting part
\begin{equation}
{\mathcal{H}}_0=\epsilon_0\hat{c}^{\dagger}\hat{c}+\sum_k\left( 
\epsilon_{kL}\hat{c}^{\dagger}_{kL}\hat{c}_{kL}+\epsilon_{kR}\hat{c}^{\dagger}_
{kR}\hat{c}_{kR}\right)+\hbar \omega_0 \hat{a}^{\dagger}\hat{a},
\end{equation}
and interacting part
\begin{equation}
{\mathcal{H}}_1=-\lambda \hat{c}^{\dagger}\hat{c}(\hat{a}+\hat{a}^{\dagger})
+\sum_k V_{kL}\left( 
\hat{c}^{\dagger}_{kL}\hat{c}+\hat{c}^{\dagger}\hat{c}_{kL}\right)
+\sum_k V_{kR}\left( 
\hat{c}^{\dagger}_{kR}\hat{c}+\hat{c}^{\dagger}\hat{c}_{kR}\right) \label{coup},
\end{equation}
where $\hat{c}$ and $\hat{a}$ operate on the states of the electron on the 
dot and the cantilever, respectively. The states in the left--hand(right--hand) 
leads have energies $\epsilon_{kL}$($\epsilon_{kR}$) and are operated on by 
$\hat{c}_{kL}$($\hat{c}_{kR}$), where the index $k$ runs over propagating 
states in the leads. The matrix elements for hopping onto(off) the dot 
are 
given by $V_{kL}$($V_{kR}$). The interaction between the electron on
the dot and the cantilever is modelled as a linear coupling between the
displacement of the flexural mode and the occupation of the dot, since
when the electron is not on the dot there is no displacement of the
cantilever.\cite{justify} We
analyze the interaction between the cantilever and the dot in some detail in 
appendix A, where the form of the interaction is derived in terms 
of the normal modes of the cantilever. We find that the coupling constant, 
$\lambda$, is given by the relation $\lambda=\xi 
eE\sqrt{\hbar/2m\omega_0}$ [see Eq.\ (\ref{coupling})], 
where $\xi$ is a geometrical factor of order one, $E$ is the electric field 
experienced by the additional electron on the dot, $m$ is the cantilever mass 
and $e (>0)$ the electronic charge.

The energy-dependent amplitude, $t_{\mathrm QD}(\epsilon)$, is calculated using
the usual  
$S$--matrix formalism\cite{W89} employed in transport theory. 
The amplitude for transmission from a state of energy $\epsilon$ to one of 
energy $\epsilon'$ is given by the element of the 
$S$--matrix\cite{W89} linking propagating states in the left and right--hand leads
\begin{equation}
\langle \epsilon', R| S|\epsilon, L \rangle =t_{\mathrm QD}(\epsilon,\epsilon'),
\end{equation} 
where $\langle  \epsilon', R|$($|  \epsilon, L\rangle$) is the state in the 
right(left)--hand lead with energy $\epsilon'$($\epsilon$). The total 
transmission amplitude at energy $\epsilon$ is then obtained by 
integrating over the final state energies
\begin{equation}
t_{\mathrm QD}(\epsilon)=\int_0^{\infty}t_{\mathrm QD}(\epsilon,\epsilon')d 
\epsilon'.\label{fermi}
\end{equation} 

The $S$--matrix element for a single electron tunneling from left to right 
through 
a single, localised, state which is coupled to a cantilever mode can be 
calculated using either the methods described by Wingreen {\emph{et al.}} 
\cite{W89} or those of Glazman and Shekhter.\cite{GS}
 For initial and final states of the cantilever
given by $\alpha_i$ 
and $\alpha_f$ respectively, the relevant matrix element is
\begin{eqnarray}
\langle \epsilon', \alpha_f, R|S|\epsilon, \alpha_i, L \rangle=&-&i\int\int 
\frac{dt_1dt_2}{\hbar^2}{\mathrm{e}}^{-\eta(|t_1|+|t_2|)}\cr
 &\times& \langle \epsilon', \alpha_f, R|
{\mathrm{e}}^{i{\mathcal{H}}_0t_2/\hbar}{\mathcal{H}}_1 
\hat{G}_R(t_2-t_1){\mathcal{H}}_1{\mathrm{e}}^{-i{\mathcal{H}}_0t_1/\hbar}
|\epsilon, \alpha_i, L \rangle,
\end{eqnarray}
where $\hat{G}_R(t)=-i\Theta(t){\mathrm{e}}^{-i{\mathcal{H}}t/\hbar}$
and $\eta$ is the usual small positive real number inserted to ensure 
convergence. 

We want to calculate the visibility of the fringes and so we need only calculate 
the coherent part of the transmission probability, and hence the elastic 
transmission amplitude. To evaluate the elastic transmission amplitude we  consider 
processes in which the state of the cantilever remains unchanged and so 
we write $\alpha_i=\alpha_f=\alpha$ and calculate an average over an ensemble 
of 
states of the cantilever (see the discussion in Sec. II above). Thus
\begin{equation}
\langle S \rangle=-V_R(\epsilon)V_L(\epsilon')\int\int 
\frac{{d}t_1dt_2}{\hbar^2}{\mathrm{e}}^{-\eta(|t_1|+|t_2|)}
\times {\mathrm{e}}^{i(\epsilon' t_2-\epsilon t_1)/\hbar}\Theta(t_2-t_1)\langle 
0,\alpha| \hat{c}(t_2)\hat{c}^{\dagger}(t_1)|0, \alpha \rangle 
\end{equation}
with 
$|V_{L(R)}(\epsilon)|^2=\sum_k|V_{kL(R)}|^2\delta(\epsilon-\epsilon_{kL(kR)})$ 
and where  $|0,\alpha\rangle$ is the state with no electrons on the 
localised level and  the cantilever in state $\alpha$.

We assume that the coupling to the leads is independent of energy over
the range of interest and that it is symmetric so that  we can write
\begin{equation}
\Gamma=\Gamma_{L(R)}=2\pi|V_{L(R)}|^2.
\end{equation}
Since we are considering only the elastic part of the transmission
amplitude the Green function will be invariant with respect to time.
 Hence, we change variables to 
$\tau=t_2-t_1$ and $t_0=t_1$, so that the Green function takes the form
\begin{equation}
G^R(\tau,t_0)=-i\Theta(\tau)\langle 
\hat{c}(\tau+t_0)\hat{c}^{\dagger}(t_0)\rangle
\end{equation}
and the  invariance with respect to translations in time is 
equivalent to  the statement that the value of the Green function is 
independent of the choice of $t_0$. Averaging over $t_0$, and taking 
the limit $\eta \rightarrow 0^+$, 
the overall expression for the $S$--matrix element is then
\begin{equation}
\langle \epsilon', R|S |\epsilon , L\rangle=
-i\frac{\Gamma}{2 
\pi}\int_0^{\infty} \frac{d\tau}{\hbar} 
{\mathrm{e}}^{i\epsilon\tau/\hbar}G^R(\tau) 
\times 2\pi\delta(\epsilon-\epsilon').
\end{equation}

Thus, the final expression for the transmission amplitude of the dot is 
\begin{equation}
\langle t_{\mathrm QD}(\epsilon)\rangle=-i\Gamma
\int_0^{\infty}\frac{d\tau}{\hbar}
{\mathrm{e}}^{i\epsilon\tau/\hbar}G^R(\tau), \label{tqd}
\end{equation}
where the transmission amplitude is understood to be time--averaged in
the sense described in Sec. II, and the 
relevant Green function is 
\begin{equation}
G^R(\tau)=-i\Theta(\tau)\langle\hat{c}(\tau)\hat{c}^{\dagger}(0)\rangle, \label{tqd22} 
\end{equation}
where the expectation value is over a thermal distribution state of the 
cantilever  with no electrons present. This result for the 
transmission amplitude  is  very similar to that
studied by Aleiner {\emph{et al.}},\cite{Al} in their analysis of the
QPC which--path device.

The Green function can be evaluated using either operator algebra
techniques,\cite{GS} or many body perturbation theory,\cite{W89} and 
one obtains
\begin{equation}
G^R(\tau)=G^R_0(\tau){\mathrm{e}}^{-\phi(\tau)}, \label{tqd2}
\end{equation}
where 
\begin{equation}
G^R_0(\tau)=-i\Theta(\tau){\mathrm{e}}^{(-i\epsilon_0-\Gamma/2)\tau/\hbar},
\end{equation}
and it has been assumed that the renormalisation of the dot energy,
$\epsilon_0$, due to coupling to the leads can be ignored.
The factor due to coupling to the cantilever, $\phi(\tau)$, can be obtained by
calculating the contribution for the cantilever beginning and ending in a
particular coherent state, $|\nu\rangle$, and then carrying out a thermal
average over all coherent states with appropriate weightings,
\begin{equation}
{\mathrm{e}}^{-\phi(\tau)}=
{\mathrm{e}}^{i(\lambda/\hbar\omega_0)^2[\omega_0\tau-\sin(\omega_0 \tau)]}
{\mathrm{e}}^{-(\lambda/\hbar\omega_0)^2[1-\cos(\omega_0\tau)]}
\int\frac{{d}^2\nu}{\pi\overline{n}}{\mathrm{e}}^
{(\lambda/\hbar\omega_0)(\nu^*\mu-\nu\mu^*)}{\mathrm{e}}
^{-|\nu|^2/\overline{n}},\label{ephi}
\end{equation}
where $\mu={\mathrm{e}}^{i\omega_0 t}-1$. 
Evaluating the thermal average, we find
\begin{equation}
\phi(\tau)
=\left(\frac{\lambda}{\hbar \omega_0}\right)^2\left\{i[\sin(\omega_0 
\tau)-\omega_0 \tau]+[1-\cos(\omega_0 \tau)](1+2\overline{n})\right\}, 
\label{ex2}
\end{equation}
with $\overline{n}$ the thermal occupation of the cantilever mode. 

It is
clear that the thermal average leads to a much more rapidly decaying
term, due to the extra factor of $2\overline{n}$. If the
cantilever remained in a coherent state throughout then it would be far less
effective, compared to the thermal state, at reducing the visibility of the 
fringes. This is because each coherent state affects the transmission amplitude 
in two different ways: the magnitude of the transmission is reduced by an amount 
which is independent of which state the cantilever is in and also
 a phase shift is induced whose size depends 
sensitively on the cantilever state.   
When we carry out the thermal average we are in effect averaging over a range
of different phase shifts. Such a procedure effectively destroys the
interference fringes whenever the thermal state includes contributions
in which the phase differs by $\sim 2\pi$, irrespective of the magnitude of the 
transmission amplitude for each of the coherent states which constitute the 
thermal mixture. The thermal averaging timescale is given by the damping time 
of the cantilever which will typically be much 
larger than the dwell time of the electron on the dot. In this case, the loss of 
fringe visibility arising from thermal smearing is  not due 
to which--path detection of individual electrons: a measurement of the  current 
averaged over times shorter than the damping time but longer than the 
dwell time would resolve  AB fringes with phase fluctuating in time.  In 
contrast, the state-independent reduction in the transmission 
amplitude would give rise to a reduction in the fringe visibility even for 
the time-resolved measurement, signifying which--path detection.   
However, since we are
 working in a steady state regime in which we consider measurements made over 
very long times we will not be able to make an explicit distinction between 
which--path detection and thermal smearing in this work.

We can repeat this calculation for the more general case of coupling
to a whole series of non-interacting cantilever modes. The result has the
same basic structure as before, but the cantilever factor, $\phi(\tau)$, 
is modified and now takes the form of a sum of the contributions from each
 mode,\cite{W89,GS}  
\begin{equation}
\phi(\tau)=\sum_i \left(\frac{\lambda_i}{\hbar \omega_i}\right)^2
\left[i(\sin(\omega_i 
\tau)-\omega_i \tau)+(1-\cos(\omega_i 
\tau))(1+2\overline{n}_i)\right],\label{phinoenv}
\end{equation}
where $\lambda_i$, $\omega_i$, and $\overline{n}_i$ are the coupling
constant, frequency, and thermal occupation number of the $i$th cantilever
mode, respectively. However, we  find that for the purposes of
demonstrating the electromechanical which--path device we need only
consider a single mode, as is discussed in appendix A.

\section{Results for Isolated Cantilever} 

The important question which we need to answer is the following:
When does the cantilever destroy the AB fringes? There are two
very different regimes which we can explore by varying the
relative sizes of characteristic dwell time of the electron on the dot, 
$\tau_d=\hbar/\Gamma$, and the fundamental frequency of the cantilever: 
\begin{enumerate}
\item $\omega_0\tau_d\ll1$: in this case the  
periodic behavior of  the cantilever will not be relevant and the entanglement
 built up between the cantilever and electron states during the dwell time will
 be irreversible, leading to dephasing. 
  
\item $\omega_0\tau_d\sim 1$: in this regime the periodicity of the
  cantilever means that the 
 entanglement of the cantilever and dot states caused by their interaction 
 may be partially undone, or erased,
 although  the distribution of electron dwell times will make the effect 
 impossible to observe directly. 
\end{enumerate}

The first regime is easier to analyze theoretically, as the
calculations can be done analytically and invite direct comparisons
with the sliding--slit  thought experiment. The case where the dwell
time of the electron on the dot is comparable to the period of the
cantilever has to be analyzed numerically. It is also complicated by
the presence of the cantilever's environment which makes the entanglement of the
cantilever and dot states irreversible by breaking the overall periodicity in the 
expression (\ref{tqd}) for the transmission amplitude. In this
section, we calculate the magnitude and phase of the transmission
amplitude in both regimes, neglecting the effect of the
environment which we address in detail in Sec. V. This approach enables
a clear picture to be built up of exactly how the environment affects the 
behavior of the cantilever--dot system.

\subsection{Regime Where $\omega_0\tau_d\ll1$}  

Since $\omega_0\tau_d\ll1$, we can simplify Eq.\
(\ref{tqd}) for
the transmission amplitude through the arm with the dot 
by expanding the harmonic functions to quadratic order in $\omega_0\tau_d$.
We obtain
\begin{equation}
 \langle t_{\mathrm QD}(\epsilon)\rangle \simeq-\Gamma 
\int_0^{\infty}\frac{{d}\tau}{\hbar}
{\mathrm{e}}^{[i(\epsilon-\epsilon_0)-\Gamma/2]\tau/\hbar
-(eE\Delta x_{\mathrm th}\tau)^2/2\hbar^2}, \label{exp}
\end{equation}
where $\Delta x_{\mathrm th}=\sqrt{(2\overline{n}+1)(\hbar/2m\omega_0)}$ is the
thermal position uncertainty of the cantilever and we have taken the 
geometrical factor in the coupling constant (\ref{coupling}) to be unity for simplicity.
The integral on the righthand side  can now be evaluated to obtain
\begin{equation}
 \langle t_{\mathrm QD}(\epsilon)\rangle \simeq -\sqrt{\frac{\pi}{2}}
\left(\frac{\Gamma}{eE\Delta x_{\mathrm th}}\right)
\exp\left(\frac{[\Gamma/2-i(\epsilon-\epsilon_0)]^2}
{2(eE\Delta {x_{\mathrm th}})^2}\right){\mathrm{Erfc}}
\left(\frac{\Gamma/2-i(\epsilon-\epsilon_0)}{\sqrt{2} eE\Delta x_{\mathrm th} }\right).
\end{equation}

If we ignore, for the moment, the thermal width of the electron energy
distribution in the leads, then we can deduce from this expression a 
rough criterion for dephasing 
in the region close to the electronic  resonance:
\begin{equation}
eE\Delta x_{\mathrm th}> \Gamma,
\label{dephasecond1}
\end{equation}
or written in another way,
\begin{equation}
eE\tau_{d}>\frac{\hbar}{\Delta x_{\mathrm th}}.
\label{dephasecond2}
\end{equation}
Practical considerations (see Sec. VI) restrict the dwell time to be 
a few ns or less and place an upper bound on the electric field: $E \sim 10^5$V/m. 
Thus, the destruction of the 
interference fringes requires
 $\Delta x_{\mathrm th}>10^{-2}$\AA, a value which is  certainly  achievable.

As previously discussed, 
the loss of fringe visibility is not only associated with which--path 
detection, but with
thermal smearing as well.  
We can obtain the condition for which--path detection 
by setting the cantilever temperature to zero in Eq. 
(\ref{dephasecond2}):
\begin{equation}
eE \tau_d >\frac{\hbar}{\Delta x_{\mathrm zp}}=2\Delta p_{\mathrm zp},
\label{whichpathcond}
\end{equation}
where $\Delta x_{\mathrm zp}=\sqrt{\hbar/2m\omega_{0}}$ denotes the 
zero-point position uncertainty. Because the reduction in the 
transmission amplitude due to each 
different coherent state is the same [c.f., Eq. (\ref{ephi})], this condition holds 
independently of which coherent state the cantilever is in.
Thus, we find that for which--path detection to occur the
classical impulse delivered to the cantilever during the dwell time must exceed
twice the zero--point momentum uncertainty of the cantilever.

Our result is  equivalent to that obtained by Bohr in his
famous discussion of Einstein's
sliding--slit {\emph{gedanken}} experiment.\cite{hist} In that case Bohr argued
that in order to
detect the passage of an electron through a given slit, the momentum 
uncertainty in the
slit must be less than the impulse transferred by the passing
electron, thus necessitating a corresponding 
latitude  in position of the slits (via the uncertainty principle)
which in turn  washes out the fringes by causing large phase shifts 
(i.e., $\sim 2\pi$) for
successive electrons. However, these phase shifts are not associated with
any kind of thermal fluctuation, instead they arise from the position uncertainty
of the quantum state of the slits.

At finite temperatures we must take into account
not only the thermal state of the cantilever, but also the fact that
 the electron energies are  spread over a range $\sim 4k_{\mathrm{B}}T$
around the Fermi energy. We must therefore average
the transmission amplitude over energy, weighted by the derivative of the
Fermi distribution function [see Eq. (\ref{therm})]:
\begin{equation}
\langle \tilde{t}_{\mathrm QD} \rangle =  \int_{0}^{\infty}{d}\epsilon
\frac{\langle t_{\mathrm QD}(\epsilon) \rangle}{4k_{\mathrm{B}}T}
{\mathrm{sech}}^2\left( \frac{\epsilon-\epsilon_0}
{2k_{\mathrm{B}}T}\right),
\end{equation}
where we have assumed that the bias across the device is small enough for a 
linear response approach to be valid and that the average Fermi energy in the 
leads is tuned to the resonance $\epsilon_0$.
The effect of this procedure is to reduce both the coherent
transmission amplitude itself and the influence the cantilever has on
it. The explanation for this unexpected feature can be found by comparing the
 transmission amplitude, $\langle t_{\mathrm QD}(\epsilon)\rangle$, at and away 
 from the 
resonance. Whilst the transmission amplitude at, or close to,  resonance 
decays rapidly with increasing 
coupling to the cantilever, the situation is reversed when the electron energy
 is far from resonance, where the transmission amplitude is {\emph{enhanced}} by
the interaction with the cantilever. The reduction in  dephasing efficiency 
due to the thermal averaging over the electron 
energy distribution makes an interesting contrast to the effect of the 
thermal average over cantilever states, which increases the dephasing rate 
substantially.

Fig.~1 illustrates the effect of increasing the dimensionless coupling constant 
$\kappa=\lambda/\hbar\omega_0$ on the magnitude of the resonant transmission 
amplitude $|\langle t_{\mathrm QD}(\epsilon_0)\rangle|$, with
and without averaging over the thermal width of the electron energy
distribution in the leads. For this example, we have taken
$\omega_0=140$~MHz and $T=$20~mK (giving a thermal occupation 
$\overline{n}=18$), $m=8\times10^{-20}$~kg, and assumed 
 $\kappa$ to have a maximum value of about 3, consistent
 with the analysis of the cantilever--dot coupling in appendix A. 
 The effect of the thermal broadening of electron energies in reducing 
 the efficiency of the cantilever to cause dephasing is clear.
 
 It is important to note  that our  
treatment is entirely restricted to a one--electron picture of transport. 
Such an approach is valid so long as the electron gas in the leads and the dot
itself is non-degenerate. When the electron temperature drops towards 
zero the interactions between electrons can no longer be ignored and so our
model and the results which follow from it become inapplicable. At 
$T=0$ and
in the weak bias limit, 
inelastic transport through the dot would become impossible: with the absence of 
unoccupied states below the Fermi level and  the cantilever in 
the ground state an electron is unable to exchange energy quanta with 
the cantilever.

\subsection{Regime Where $\omega_0\tau_d\sim 1$}           

In the regime where the dwell time of the electron on the dot
approaches the period of the cantilever, the system takes on
a rather different character, as the dephasing interaction between the
cantilever and the electron on the dot is no longer a brief scattering
event, but a sustained interaction. In this regime we would 
expect to see evidence of the periodic behavior of the cantilever,
such as side resonances in the elastic transmission amplitude, but
in practice these would have to compete with thermal effects which tend to wash out
fine structure in the transmission characteristics.

In order to calculate the transmission amplitude, 
it is convenient to recast our earlier expression (\ref{tqd}) 
in the form
\begin{eqnarray}
\langle t_{\mathrm QD}(\epsilon) \rangle&=& -\frac{\Gamma}{\hbar\omega_0}
\frac{{\mathrm{e}}^{-(\lambda/\hbar\omega_0)^2(2\overline{n}+1)}}
{1-{\mathrm{e}}^{-(2\pi/\omega_0)[i(\epsilon_0-(\lambda^2/\hbar\omega_0)
-\epsilon)/\hbar)+\Gamma/2\hbar]}}
\nonumber \\ 
&&\times \int_0^{2\pi}{d}\tau'
\exp\Biggl\{\frac{-\tau'}{\omega_0}\left[\frac{i}{\hbar}
\left(\epsilon_0-\frac{\lambda^2}{\hbar\omega_0}-\epsilon \right)
+\frac{\Gamma}{2\hbar}  \right]\nonumber \\       
&&-i\left(\frac{\lambda}{\hbar\omega_0}\right)^2 \sin (\tau')
+(2\overline{n}+1)\left(\frac{\lambda}{\hbar\omega_0}\right)^2 
\cos (\tau')\Biggr\},\label{calc}
\end{eqnarray}
where $\tau'=\omega_0 \tau$. Of course, to model a practical experiment, we also 
need to average over the energy, $\epsilon$, to take into
account the effect of electron temperature. However, we
 look first at the dependence of the transmission amplitude on the incident energy. 
 Although this  would require `monochromatic' 
electrons,  important insight is gained into the behavior of the system
which in practice may be obscured by thermal effects.

In Fig.~2, the magnitude of the resonant transmission amplitude is plotted
against the coupling $\kappa$ for $\Gamma/\hbar\omega_0$=2, 1, and
 0.5, and with $\overline{n}=18$. The behavior is similar in all three cases for 
 small $\kappa$, with a rapid decay in the magnitude of the transmission amplitude,
  due to dephasing caused by a combination of  which--path detection and
 thermal fluctuations in the state of the cantilever.
  Notice, however, that
 for $\Gamma/\hbar\omega_0$=1,  oscillations eventually develop in the 
transmission, and that these become even more pronounced 
for $\Gamma/\hbar\omega_0$=0.5.

In order to clarify the origin of the oscillations in
 $|\langle t_{\mathrm QD}(\epsilon)\rangle|$,
 Fig.~3 plots the evolution of the amplitude and phase of the
transmission amplitude as functions of both the coupling constant,
$\kappa$, and the detuning energy $\epsilon-\epsilon_0$, for 
$\Gamma/\hbar \omega_0=0.5$ and $\overline{n}=18$.  
It is now clear that there are side resonances at
$\epsilon-\epsilon_{0}+\lambda^2/\hbar\omega_0=
\pm\hbar\omega_0,\pm2\hbar\omega_0,...$, and that
these resonances drift in energy as the value of $\kappa$
is increased.\cite{W89,JW} 
It is this drift  which is responsible for the oscillations
 in $|\langle t_{\mathrm QD}(\epsilon)\rangle|$ as a function of the coupling, $\kappa$, 
for given fixed $\epsilon$. 

Note that if the cantilever temperature is set equal to 
zero, i.e., it is in 
its ground state, 
$\overline{n}=0$, then side resonances are still observed, but only on one 
side of the main peak, at $\epsilon-\epsilon_{0}+\lambda^2/\hbar\omega_0
=+\hbar\omega_0,
+2\hbar\omega_0,...$.This is because in its ground state the cantilever  
can only absorb energy quanta.

Under the conditions $k_{\mathrm B}T\gg\hbar\omega_{0}$, 
$\kappa\sim 1$, and near resonance, Eq. 
(\ref{calc}) has the following asymptotic approximation:
\begin{equation}
    \langle t_{\mathrm QD}(\epsilon) \rangle 
    \sim 
    -\frac{\Gamma}{2\lambda}\sqrt{\frac{\pi\hbar\omega_{0}}{k_{\mathrm 
    B}T}}
    \left\{\frac
{1+{\mathrm{e}}^{-(2\pi/\hbar\omega_0)[i(\epsilon_0-\lambda^2/\hbar\omega_0
-\epsilon)+\Gamma/2\hbar]}}{1-{\mathrm{e}}^{-(2\pi/\hbar\omega_0)
[i(\epsilon_0-\lambda^2/\hbar\omega_0
-\epsilon)+\Gamma/2\hbar]}}\right\}.\label{asympt}
\end{equation} 
The various above-discussed features in the transmission 
amplitude dependence on coupling and energy can be clearly seen in 
this simplifying approximation.

Side resonances are a familiar feature in tunneling problems, but unlike those
found here they are usually 
associated with inelastic processes. However, as was  first
shown by Jauho and Wingreen,\cite{JW} who considered the classical--oscillator
version of the problem we address, such resonances can also occur in the
elastic transmission amplitude.  Side resonances in the elastic transmission 
indicate
coherent or virtual exchange of energy quanta between the electron on the dot 
and the cantilever, with no net energy interchange over the dwell 
time of the electron.  
In a fully quantum mechanical system, these processes must rely on the coupled 
cantilever--electron system maintaining its phase coherency and so we may expect 
that the influence on the cantilever of its environment should be detectable, 
at least in principle,  via its effect on these resonances. 

The side resonances have a separation in energy of $\hbar\omega_0$, but  
for any practically realizable system the thermal energy scale will be 
much larger, as we discuss in detail in Sec. VI. However, in  
the regime where $\hbar\omega_0\ll k_{\mathrm{B}} T$,
averaging over the thermal distribution of electron energies will wash 
out the side resonances.
 This means that it would not be
possible to find any trace of the coherent exchange of energy quanta between
the cantilever and the electron on the dot in the transmission 
characteristics. Only if the
electron energy width could be lowered or the frequency raised to the
point where $\hbar\omega_0\sim k_{\mathrm{B}}T$ would we then expect
such features to be visible. Note, in the case of photon assisted tunneling the photon
energy can be made comparable with $k_{\mathrm{B}} T$  
without difficulty.\cite{JW}

\section{Influence of Environment on Cantilever}
Thus far, we have treated the cantilever--dot system as isolated,
thereby ignoring the effect of the environment on their degrees of
freedom. This approximation is valid when the dwell time of the electron on 
the dot is  short compared to the decoherence time of the 
cantilever, 
as will most certainly be the case for $\omega_0\tau_d\ll1$. However, there is no 
reason  to assume
that this will also be the case when $\omega_0\tau_d\sim1$.  Indeed, we
would like to know how the environment of the cantilever affects the coherent 
oscillations which occur in the transmission amplitude as a function of the 
coupling strength  (for monochromatic electrons). Intuitively, we expect that the 
decoherence arising from the cantilever's environment  should wash out the side 
resonances in the transmission amplitude, but a detailed calculation is required 
to determine  the cantilever quality ($Q$) factor range
 for this to 
occur.

In general,
both the interaction between the dot electron and its local environment (other
electrons, phonons, photons, etc. in the dot region) and the interaction
between the cantilever and its environment (the other collective vibrational modes,
as well as internal electronic degrees of freedom, external photons, 
gas molecules, etc.) will
contribute to the decoherence. However, since we can always measure the
properties of the fringes with the dot--cantilever interaction turned
off for an arbitrary dwell time ($\tau_d=\hbar/\Gamma$), we need only consider
the additional effect of the cantilever's environment, as the electron's environment
can be included via a renormalisation of the zero electric field
transmission amplitude.

\subsection{Estimate of Cantilever Decoherence Time}

We can obtain a rough estimate of the time--scale over which superpositions of 
cantilever states, resulting from the coupling to the dot electron, 
are likely to be decohered by the environment by using
a simple, heuristic approach which models the environment as an infinite set of 
harmonic oscillators.\cite{UZ,Bose} This 
approach leads to the prediction that for
a system with  a classical damping rate, $\gamma_c$, a linear superposition of
 two different coherent
states whose centers are a distance $\Delta x$ apart, where $\Delta x$
is greater than the thermal
de Broglie wavelength, $\lambda_{\mathrm th}=\hbar/\sqrt{2m k_{\mathrm{B}}T}$,
will decohere at a rate $\gamma_d$, given by
\begin{equation}
\gamma_d\sim\frac{2m\gamma_c k_{\mathrm{B}}T(\Delta x)^2}{\hbar^2}.\label{rot}
\end{equation}

However, we cannot apply this heuristic rule directly to the coupled 
cantilever--dot system as the displacement of the cantilever is a continuously 
varying function. An estimate for the decoherence rate of the cantilever can 
only be obtained using this method if the further approximation of 
averaging over the cantilever displacement is made.\cite{Bose} 
If the cantilever starts in any given 
coherent state when the electron arrives on the dot, then the
state of the cantilever at time $t$ later will be a different coherent state
centered a distance $\Delta x$ apart from the point on which the initial 
state was centred, where 
\begin{equation}
\Delta x(t)=\sqrt{\frac{2\hbar}{m\omega_0}}\ 
\kappa[1-\cos(\omega_0 t)].
\end{equation}
Thus we can obtain an estimate for the decoherence rate $\gamma_{d}$ of the 
cantilever due to its
environment for the case where $\tau_d\omega_0=1$ by averaging this displacement 
over the cantilever period:\cite{Bose}
\begin{equation}
\gamma_d \sim\frac{1}{(2\pi/\omega_0)}\frac{4\kappa^2\gamma_c k_{\mathrm{B}}T}
{\hbar\omega_0}\int_0^{2\pi/\omega_0}{d}t[1-\cos(\omega_0 
t)]^2
=6\kappa^2\gamma_c\frac{k_{\mathrm{B}}T}{\hbar 
\omega_0}.\label{decoherence}
\end{equation} 
For a nanotube cantilever of frequency $\sim140$MHz (see appendix A), 
the $Q$--factor\cite{vib}
can be of order $500$ and so the classical damping rate, $\gamma_c$, will be 
of order $3\times 10^5$~$\mathrm{s}^{-1}$. Thus, at a temperature of 
20~mK, our
heuristic expression for the decoherence rate of the nanotube cantilever  gives 
$1/\gamma_d\sim 3\times 10^{-9}$~s for $\kappa\simeq 3$. This result 
signals that  the decoherence of the cantilever due to its environment has an effect
whenever the dwell time approaches a magnitude of order $1$ns.

Whilst the heuristic model we have outlined is very useful for
estimating whether or not the cantilever's environment is relevant for
the calculation of the transmission amplitude, it is not expected to
be very accurate, as it is an approximation even of the rule--of--thumb given 
by   Eq. (\ref{rot}).
In order to improve on this estimate, 
we must  enlarge our simple model by adding to the system 
Hamiltonian (\ref{Ham1}) the cantilever's 
environment, modelled as an infinite bath of harmonic oscillators with 
linear coupling to the cantilever.  
Modelling the environment in this way is of course itself a fairly
serious approximation. However, this approximation is ubiquitous in
one form or another throughout the theory of open quantum 
systems.\cite{CL,UZ,O,JZ} By extending our calculation to include this
 model of the cantilever's environment, we can obtain predictions
for how the transmission properties of our which--path device depend
on the decoherence rate of the cantilever. Thus, we can use our
theory to predict how changing the $Q$--factor of the cantilever
affects the interference fringes.         

\subsection{Transport Properties with Environmental Coupling}

The standard model of the environment which we use is  an infinite bath of harmonic 
oscillators which interact linearly with the cantilever, but do not interact 
with each other. We have to assume an infinite bath of oscillators in order to 
have a reservoir which remains in thermal equilibrium despite contact with the 
cantilever. The infinite oscillator--bath model of the environment is also 
equivalent to a quantum Langevin formalism in which the cantilever operators 
have an equation of motion containing a damping term and a thermal noise 
operator arising from the reservoir. 

We begin by considering just the interaction between the cantilever and its
environment, before going on to include the effect of an electron on the dot and
hence obtaining the transmission amplitude including the environment.
The cantilever-environment Hamiltonian can be written as\cite{MW,Lou}
\begin{equation}
{\mathcal{H}}^{c}=\hbar \omega_0\hat{a}^{\dagger}\hat{a}
+\sum_{\omega} \hbar \omega 
\hat{A}^{\dagger}(\omega)\hat{A}(\omega) 
+\sum_{\omega}[g(\omega)\hat{a}^{\dagger}\hat{A}(\omega) + 
g^*(\omega)\hat{A}^{\dagger}(\omega)\hat{a}], \label{envham}
\end{equation}
where the cantilever states are again operated on by 
$\hat{a}$, the bath oscillator of frequency $\omega$ is operated on by
  $\hat{A}(\omega)$ and the coupling constants $g(\omega)$ depend on
  the bath oscillator frequency.

 Using the Hamiltonian ${\mathcal{H}}^c$, the  equation 
of motion for $\hat{a}$ in the interaction picture is 
readily obtained,\cite{MW}
\begin{equation}
\dot{\hat{a}}(t)=(-i \omega-\gamma)\hat{a}(t)-\hat{F}(t),
\end{equation}
where 
\begin{equation}
\hat{F}(t)=i\sum_\omega g(\omega)\hat{A}(\omega,0){\mathrm{e}}^{-i \omega t}
\end{equation}
and the damping coefficient, $\gamma$, is given by
\begin{equation}
\gamma=\pi \eta_{b}(\omega_0)|g(\omega_0)|^2,
\end{equation}
where $\eta_{b}(\omega)\delta \omega$ is the number of bath oscillators in the 
spectral range $\delta\omega$. The coefficient $\gamma$ provides the bridge 
between the model and experiment, as it is simply related to the rate at which the 
system loses energy after being excited (i.e., the classical damping rate),\cite{Lou}
$\gamma_c=2\gamma$.

Integration of the equation of motion for $\hat{a}(t)$, and a similar one for 
$\hat{a}^{\dagger}(t)$, leads to explicit expressions for 
$\hat{a}(t)$ and $\hat{a}^{\dagger}(t)$:
\begin{equation}
\hat{a}(t)=\hat{a}(0){\mathrm{e}}^{(-i\omega_0-\gamma)t}-{\mathrm{e}}
^{(-i\omega_0- 
\gamma)t}\int_0^t dt' \hat{F}(t'){\mathrm{e}}^{(i\omega_0+\gamma)t'}, 
\label{plaina}
\end{equation} 
\begin{equation}
\hat{a}^{\dagger}(t)=\hat{a}^{\dagger}(0){\mathrm{e}}^{(i\omega_0-\gamma)t}-
{\mathrm{e}}^{(i 
\omega_0-\gamma)t}\int_0^t dt' 
\hat{F}^{\dagger}(t'){\mathrm{e}}^{(-i\omega_0+\gamma)t'}. \label{adag}
\end{equation}  

The time $t=0$ holds a special place in the theory since it is the time  at 
which the interactions between the cantilever and the bath are turned on and 
so the cantilever and the bath are apparently quite independent at this instant.
 However, this  need not be the case and certainly would not be appropriate 
 for the system  we are considering. In
order to specify the model completely we need to give the expectation values 
at $t=0$ for both the cantilever and the bath. If we set the initial 
expectation values of the cantilever to be those of a thermal state at the
same temperature as the bath, then we can describe the case in which the
cantilever has been interacting with the bath for a time much longer than the 
damping time, $1/\gamma_c$, {\emph{before}} $t=0$ and so is in equilibrium with
the bath. Our only assumption is that
quantum correlations between the cantilever and the bath can be neglected
at $t=0$, the usual assumption made in calculating decoherence rates.

 The expectation values are over  
products of  cantilever and environment states.  We eventually want to calculate the 
transmission amplitude for an electron interacting with
a cantilever which is initially in a thermal state
and so we choose to define the $t=0$ expectation values as:
\begin{equation}
\langle...\rangle=\frac{{\mathrm{Tr}}[...{\mathrm{e}}
^{-{\mathcal{H}}^c_0/k_{\mathrm{B}}T}]}
{{\mathrm{Tr}}[{\mathrm{e}}^{-{{\mathcal{H}}^c_0}/k_{\mathrm{B}}T}]},
\end{equation}
where the Hamiltonian without interactions is 
\begin{equation}
{\mathcal{H}}^c_0=\hbar \omega_0\hat{a}^{\dagger}\hat{a}
+\sum_{\omega} \hbar \omega 
\hat{A}^{\dagger}(\omega)\hat{A}(\omega),
\end{equation}
and $T$ defines the fixed temperature of the environment.
Using this definition we find
\begin{eqnarray}
\langle \hat{A}(\omega,0)\rangle&=&0 \nonumber \\
\langle \hat{A}(\omega,0)\hat{A}(\omega',0)\rangle&=&0 \\
\langle \hat{A}^{\dagger}(\omega,0)\hat{A}(\omega',0)\rangle&=&\delta_{\omega 
\omega'}N(\omega),\nonumber 
\end{eqnarray}   
with
\begin{equation}
N(\omega)=\frac{1}{{{\mathrm{e}}}^{\hbar \omega/k_{\mathrm{B}}T}-1}.
\end{equation}

For the cantilever itself the values of the zero-time correlation functions 
represent the initial conditions of the problem. In 
thermal equilibrium with the environment at temperature $T$,   
\begin{eqnarray}
\langle \hat{a}(0)\rangle&=&0\nonumber \\
\langle \hat{a}(0)\hat{a}(0)\rangle&=&0 \\
\langle \hat{a}^{\dagger}(0)\hat{a}(0)\rangle&=&\overline{n},\nonumber
\end{eqnarray}   
with
\begin{equation}
\overline{n}\equiv N(\omega_0)=\frac{1}{{\mathrm{e}}^{\hbar 
\omega_0/k_{\mathrm{B}}T}-1}.
\end{equation}

The purpose of extending the analysis to include the interactions between the
cantilever and its environment is to see how they modify the 
interaction between the cantilever and the electron {\emph{whilst it is on the dot}}. 
Therefore we can choose our
origin of  time, and hence the definition of the initial cantilever state, to 
be the time when the electron jumps onto the dot. There is no need to explicitly
include the interaction with the bath before the electron hops onto the dot as it is
already implicitly included by assuming the cantilever is in a thermal state 
at temperature $T$. However, when the electron is on the dot it drives 
the cantilever away from equilibrium and we now need to include the 
interaction with the environment explicitly to model the behavior 
of the cantilever--dot system as accurately as possible.

We obtain the transmission amplitude for the dot, including the 
effect of the cantilever's environment by applying the $S$--matrix we used in
 Sec. III to a generalization of the Hamiltonian (\ref{Ham1})
 which includes the coupling of the cantilever to 
 the bath of oscillators,
\begin{equation}
{\mathcal{H}}=\epsilon_0\hat{c}^{\dagger}\hat{c}+{\mathcal{H}}^c
+{\mathcal{H}}_1+\sum_k\left( 
\epsilon_{kL}\hat{c}^{\dagger}_{kL}\hat{c}_{kL}+\epsilon_{kR}\hat{c}^{\dagger}_
{kR}\hat{c}_{kR}\right), 
\end{equation}
where ${\mathcal{H}}_1$, given by eqn (\ref{coup}), describes the 
electron--cantilever coupling and the cantilever--environment part, ${\mathcal{H}}^c$, 
is given by Eq.\ (\ref{envham}).

Using the methods of Glazman and Shekhter,\cite{GS} we can again separate out 
the electronic 
part of the transmission amplitude from the average over cantilever states, 
and so we find
\begin{eqnarray}
\langle t_{\mathrm QD}(\epsilon) \rangle &=& 
-\Gamma\int_0^{\infty}\frac{dt}{\hbar}{\mathrm{e}}^{[i(\epsilon-\epsilon_0)
-\Gamma/2]t/\hbar}
\langle{\mathrm{T}}_t 
{\mathrm{e}}^{-i \int_0^{t}W_I(t')dt'/\hbar} \rangle.
\end{eqnarray}
The term in angled brackets on the right--hand side is known as the influence
 functional, 
${\mathrm{T}}_t$ is the time ordering operator, and $W_I(t)$  is the
 electron--cantilever coupling defined in the interaction picture 
\begin{equation}
W_I(t)=-{\mathrm{e}}^{i{\mathcal{H}}^c t/\hbar}
\lambda(\hat{a}^{\dagger} +\hat{a})
{\mathrm{e}}^{-i{\mathcal{H}}^c t/\hbar}
=-\lambda[ \hat{a}^{\dagger}(t) +\hat{a}(t)],
\end{equation}
where $\hat{a}(t)$ and $\hat{a}^{\dagger}(t)$ are given by Eqs. (\ref{plaina}) and 
(\ref{adag}) above.

Here, we shall
carry out a much simpler, approximate calculation of the influence 
functional which involves performing a second order expansion in 
$\lambda$ and then re-exponentiating. In appendix B, it is shown that 
this approximate calculation in fact coincides with the exact result 
obtained from a full calculation. 

The expansion to second order gives 
\begin{equation}
\langle{\mathrm{T}}_t {\mathrm{e}}^{-i \int_0^{t}W_I(t')dt'/\hbar} \rangle=
1-i\int_0^t \frac{dt'}{\hbar}\langle W(t')\rangle 
+(-i)^2\int_0^t \frac{dt'}{\hbar}\int_0^{t'}\frac{dt''}{\hbar}\langle 
W(t')W(t'')\rangle. \label{if}
\end{equation}  
The next step is to evaluate the correlation 
functions which arise in the first and second order terms in the 
expansion of the influence functional. At first order
\begin{equation}
\langle W(t') \rangle=-\lambda\langle  [\hat{a}(t')+\hat{a}^{\dagger}(t')]\rangle,
\end{equation}
 and at second order
\begin{equation}
\langle W(t')W(t'')\rangle = \lambda^2\{ \langle 
\hat{a}(t') \hat{a}(t'')\rangle+ \langle \hat{a}(t') 
\hat{a}^{\dagger}(t'')\rangle   
+\langle \hat{a}^{\dagger}(t') \hat{a}(t'')\rangle + \langle 
\hat{a}^{\dagger}(t') \hat{a}^{\dagger}(t'')\rangle \},
\end{equation} 
where $t'\geq t''$.

We can calculate all the correlation functions 
 using the initial conditions defined above and the standard results of 
quantum Langevin theory.\cite{MW,Lou}
We  start by observing that three of the five correlation functions are 
zero:
\begin{equation}
\langle \hat{a}(t')\hat{a}(t'')\rangle=\langle 
\hat{a}^{\dagger}(t')\hat{a}^{\dagger}(t'')\rangle=0 \label{a}
\end{equation}
and
\begin{equation}
\langle [\hat{a}(t')+\hat{a}^{\dagger}(t')]\rangle=0.
\end{equation}
These results follow from the definitions above, since correlators of the type 
$\langle \hat{a}(0)\hat{F}(t) \rangle$  decouple into products of one-time 
correlation functions and the definitions imply that $\langle \hat{F}(t) 
\rangle=\langle \hat{F}(t')\hat{F}(t'') \rangle=0$. 
For the other two correlation functions, we have
\begin{eqnarray}
\langle \hat{a}(t')\hat{a}^{\dagger}(t'')\rangle&=&  (\overline{n}+1)
{\mathrm{e}}^{-i\omega_0 (t'-t'')} {\mathrm{e}}^{-\gamma (t'-t'')} 
\label{b} \\
\langle \hat{a}^{\dagger}(t')\hat{a}(t'')\rangle&=&  \overline{n}
{\mathrm{e}}^{i\omega_0 (t'-t'')} {\mathrm{e}}^{-\gamma (t'-t'')} \label{c}.
\end{eqnarray}
Thus, we again find
\begin{equation}
\langle t_{\mathrm QD}(\epsilon) \rangle=-\Gamma\int_0^{\infty}\frac{dt}{\hbar}
{\mathrm{e}}^{[i(\epsilon-\epsilon_0)
-\Gamma/2]t/\hbar}{\mathrm{e}}^{-\phi(t)}, 
\label{exact}
\end{equation} 
but now with 
\begin{equation}
\phi(t)=\frac{\lambda^2}{\hbar^2} \int_0^t 
{dt'}\int^{t'}_{0}{dt''}{{\mathrm{e}}}^{-\gamma (t'-t'')}\left[ (\overline{n}+1)
{\mathrm{e}}^{-i\omega_0 (t'-t'')}+\overline{n}
{\mathrm{e}}^{i\omega_0 (t'-t'')}\right], 
\label{phi}
\end{equation}
where we have re-exponentiated the expansion in $\lambda$.

Carrying out a change of variables to $\tau=t'-t''$, the integrals are readily 
evaluated to give
\begin{equation}
\phi(t)=\left(\frac{\lambda}{\hbar}\right)^2\left[ 
\frac{(\overline{n}+1)}{(\gamma+i\omega_0)}\left(t+\frac{{\mathrm{e}}^{-(\gamma+i
\omega_0)t}-1}{\gamma+i\omega_0}\right)+\frac{\overline{n}}{(\gamma-i\omega_0)}
\left(t+\frac{{\mathrm{e}}^{-(\gamma-i\omega_0)t}-1}{\gamma-i\omega_0}\right)
\right]. \label{phix}
\end{equation}
Setting $\gamma=0$, one may verify that (\ref{phix}) indeed reduces 
to expression (\ref{ex2}) in the absence of the environment.

Under the conditions $Q\gg 1$, $k_{\mathrm B}T\gg\hbar\omega_{0}$, 
$\kappa\sim 1$, and near resonance, the transmission amplitude 
(\ref{exact}) has the following asymptotic approximation:
\begin{equation}
    \langle t_{\mathrm QD}(\epsilon) \rangle 
    \sim 
    -\frac{\Gamma}{2\lambda}\sqrt{\frac{\pi\hbar\omega_{0}}{k_{\mathrm 
    B}T}}
    \left\{\frac
{1+{\mathrm{e}}^{-(2\pi/\hbar\omega_0)[i(\epsilon_0-\lambda^2/\hbar\omega_0
-\epsilon)+\Gamma/2\hbar+(\kappa^{2}\gamma_{c}k_{\mathrm 
B}T/\hbar\omega_{0})]}}{1-{\mathrm{e}}^{-(2\pi/\hbar\omega_0)
[i(\epsilon_0-\lambda^2/\hbar\omega_0
-\epsilon)+\Gamma/2\hbar+(\kappa^{2}\gamma_{c}k_{\mathrm 
B}T/\hbar\omega_{0})]}}\right\}.\label{asymptenv}
\end{equation} 
Note that this approximation differs from the earlier-derived one for 
the transmission amplitude 
in the absence of the environment, Eq. (\ref{asympt}), merely by the 
replacement of $\Gamma/2\hbar$ with $\Gamma/2\hbar+(\kappa^{2}\gamma_{c}k_{\mathrm 
B}T/\hbar\omega_{0})$. Thus, a decoherence rate $\gamma_{d}$ can be identified 
as the term $\kappa^{2}\gamma_{c}k_{\mathrm B}T/\hbar\omega_{0}$, which 
agrees with the earlier-derived estimate 
(\ref{decoherence}) up to an overall numerical factor. 
Approximation (\ref{asymptenv}) clearly shows the washing out of
the coherent, oscillatory behavior by the 
environment when the decoherence time $1/\gamma_{d}$ is shorter than 
the cantilever period. When the decoherence time exceeds the dwell 
time $\hbar/\Gamma$, then the former has a 
negligible effect on the transmission properties.

\subsection{Results for Cantilever Coupled to Environment}
The expression for the transmission amplitude including the coupling to
the cantilever's environment [Eq.\ (\ref{exact})] can be integrated
numerically, thereby allowing us to explore the effect on the
transmission characteristics of varying the cantilever's
$Q$--factor under more general 
conditions than those for which approximation (\ref{asymptenv}) is 
justified. Fig.~4 shows the magnitude 
and phase of the
transmission  amplitude through the dot for $\Gamma/\hbar \omega_0=0.5$,
$\overline{n}=18$ and $Q=\omega_0/\gamma_{c}=50$. The diagram 
should be compared with
Fig.~3 showing the behavior of the same system without
environmental coupling. Fig.~5 shows the behavior at resonance of
$|\langle t_{\mathrm QD}(\epsilon)\rangle|$ for 
$Q=50$ and 500, as well as the $\gamma_{c}/\omega_0 =0$  (no environment)
case for comparison.

It is clear from the figures that the coupling to the environment tends to
destroy the side resonances in the transmission amplitude, as well as
the associated features in the phase. This is because the environment acts to 
degrade the coherent 
superposition of cantilever states into which the interaction with the electron tries
to drive the cantilever. Furthermore, the figures show that these 
 environmental effects 
become increasingly important as the cantilever--electron coupling,
$\kappa$, is increased. This is because the larger $\kappa$ is, the greater 
is the 
 difference between the 
states in the superposition into which the cantilever is driven, and consequently 
the faster is the rate at which the superposition decoheres.

\section{Practical Considerations}

The model parameter ranges actually allowed in an experiment  are limited by 
practical constraints. Up until now we 
have only referred to these very loosely and have concentrated instead on the
range of behavior which can occur in the which--path system under fundamental 
constraints alone.

Probably
the most important practical constraint affects the upper range of the electric field
which can be developed between the cantilever and the dot. 
The maximum allowable field is typically $\sim 10^5$~V/m before 
breakdown of the two--dimensional electron gas occurs due to  
deconfinement,\cite{Buksp} and so we
must take this as the largest possible value in considering the
practicality of the system. 

The temperature of
 the system is limited by the difficulty of cooling conduction electrons  
 to ultralow temperatures. It becomes extremely difficult to reduce the 
 electron temperature  below about 20~mK,
 because the electrons become practically decoupled from acoustic 
 phonons. 
 Therefore we take 20~mK to be 
 the temperature minimum. 

The frequency range of the cantilever is crucial 
for observing quantum coherent behavior such as  side resonances, as 
well as their destruction 
due to 
decoherence. The lower limit 
on the frequency range is set by the requirement that the cantilever 
period should be comparable to the dwell time $\tau_d=\hbar/\Gamma$ of the electron on the quantum dot,  
while the 
upper limit is set
 by the requirement that the interaction between the cantilever and
the electron on the dot be sufficiently strong, given the limits on
the electric field,  to lead to decoherence,
again on a time--scale  comparable to the dwell time.
The interaction between
the electron on the dot and the cantilever is discussed in more
detail in appendix A, along with the geometrical factors which arise in
the calculation of the effect of the electric field on the flexural cantilever
modes.
The dwell time  is
limited by the rate at which processes other than the interaction 
with the cantilever cause decoherence, as well as how small a current can be measured 
through the dot. Of course, the decoherence rate
of all the background processes is very difficult to estimate
theoretically; in experiment it can be done by
observing the visibility of the fringes as a function of dwell time
with the cantilever interaction switched off. This then provides a
baseline with which to compare all later measurements where we are
interested in the effect of the cantilever. Previous experiments 
carried out at $T=0.1$~K by
Yacoby {\emph{et al.}},\cite{Yac} which were designed to measure the phase of
the transmission amplitude through a dot in the absence of any
external probe, show that dwell times as long 10~ns and currents as low as 
$\sim10^{-11}$~A lead to fringes which are still detectable.

It turns out that the best compromise between the two 
competing frequency limits is achieved by using say a carbon nanotube 
cantilever with a frequency of 100--200~MHz, giving a maximum 
coupling $\kappa=\lambda/\hbar\omega_{0}\sim 3$ (appendix A). The   
problem is not so much in finding a cantilever with a high enough frequency, but rather
in obtaining a large enough coupling from the electric field to cause a 
detectable amount of dephasing.

In the light of these practical constraints it is clear that only some 
 of the theoretical results obtained in our
analysis would be observable in an experiment using current
 technology. The overall destruction of the
 AB interference fringes as the coupling between the cantilever and the
 electron on the dot is increased should be detectable, both in the
 regime where $\Gamma/\hbar\omega_0\gg 1$ and $\Gamma/\hbar\omega_0\sim1$.
However, the restrictions on the temperature and cantilever
 frequency imply that an experiment would have to be performed in the regime
 where $k_{\mathrm{B}}T\gg \hbar\omega_0$. This means that the thermal
 width of electron energies will wash out the side resonances 
 in the transmission which were found to occur when 
$\Gamma/\hbar\omega_0\sim1$. Under these circumstances, the effect of
 the cantilever's environment on the transmission characteristics
 would be obscured. However, we emphasise again that these
 limitations are not fundamental: if the width of the electron
 energies could be reduced or if the restriction on the maximum
 cantilever frequency could be relaxed (by finding a way to increase
 the coupling between the electron on the dot and the cantilever, for
 example) then the coherent, quantum electromechanical features which our
 analysis predicts should be observable. 
 Whilst it is not clear how
 the coupling constant between the cantilever and the dot could be
 increased, it is possible to reduce the thermal width of
 electron energies by using a double quantum dot system 
 rather than a single one as we have considered here.\cite{Buks2}  
   
It is also  possible to conduct experiments in which  an ac bias is applied 
to the AB ring. Such an experiment would provide an alternative  way of 
investigating
the interaction between the cantilever and its environment: if the ac frequency 
is higher than the rate $\gamma_{c}$ at which the cantilever state changes due to 
thermal fluctuations then the thermally-induced fluctuations in the phase of the 
interference fringes should  be detectable, and distinguishable from 
the destruction of the interference fringes due to which-path 
detection. However, a detailed analysis of such
an experiment requires extending the theory developed here to include
time dependence, and so goes well beyond the present analysis.

\section{Conclusions and Discussion}

We have carried out a theoretical analysis of a possible solid state
which--path interferometer, in which electronic and mechanical degrees of freedom 
become coupled. The visibility and phase of the interference
fringes in the system depend strongly on the coupling between the
electron on the dot and the cantilever. The reduction in visibility 
with increasing coupling is 
due in part to
the cantilever measuring the path taken by the electron and also due 
to thermal
fluctuations in the state of the cantilever.

When the dwell time of the electron on the dot is short compared to
the period of the cantilever, the system behaves in a way which is
analogous to Einstein's celebrated recoiling slit experiment. In
contrast, when the dwell time is comparable to the cantilever's period,
 the cantilever and the electron on the
dot   show signs of behaving  as a single coherent quantum system,
 so long as the electrons incident on the dot have a
sufficiently narrow energy width.  The coherency of the
cantilever--dot system is inferred from the appearance of side
resonances in the transmission characteristics. Including  the cantilever's
 environment in the analysis, we find that the side
 resonances are washed out   for small enough  oscillator 
 $Q$--factor, while the average decrease in the fringe visibility with
 increasing coupling to the cantilever is not affected.

The basic feature of the reduction of fringe visibility as a function
of the coupling between the electron on the dot and the cantilever
should be observable in an experiment using currently available technology. 
However, the 
more delicate features   such as the
side resonances in the transmission amplitude for long dwell times 
and the effect of the 
 cantilever's environment on these resonances will usually be
 obscured by the  energy thermal width of electrons incident on the dot.

There are two important ways in which our analysis could be
extended. Most straightforwardly, we could examine what effect using a
double dot, rather than a single one, would have on the behavior of the
system. In particular, it would be interesting to see to what extent the
electron energy width could be reduced, whilst maintaing
an overall, measurable current through the device. 
The second way our
analysis could be extended would be to go beyond our steady state
treatment to obtain the time dependence of the transmission
amplitude. A time dependent analysis would allow us to make predictions
about the way in
which the thermal fluctuations in the state of the cantilever cause
fluctuations in the phase of the interference fringes. Such
fluctuations may prove to be observable in ac experiments and may also
provide us with another way of inferring information about how the
cantilever interacts with its environment.
  
\acknowledgements
M.P.B. would especially like to thank Eyal Buks for many helpful 
conversations and invaluable suggestions. Very helpful conversations 
with Sougato Bose are also gratefully acknowledged.
A.D.A. thanks Angus MacKinnon for a series of very useful discussions.  Funding was provided by the 
EPSRC under Grant No. GR/M42909/01.

\appendix
\section{Cantilever}

The cantilever in the which--path experiment we propose must fulfil a
number of quite stringent properties: it must have a fundamental 
frequency in the range 100--200~MHz and it must also  
be a conductor. These requirements can be satisfied conveniently
 by using a carbon nanotube as a cantilever, rather than a device which has
been fabricated via some kind of etching process from a much larger substrate. 

Carbon nanotubes have a number of remarkable physical properties which
are beginning to exploited for practical purposes. Recent experiments
have seen them employed as hyper--sensitive tips in AFM experiments,
 with the nanotube attached to the end of a conventional AFM tip to
 extend the effective range of resolution of the device.\cite{AFM} A similar
 apparatus could be used to bring a nanotube into position to act as
 the cantilever in our which--path experiment, either in the
 geometrical configuration explicitly considered above or some
 variation of it  which would lead to slightly different geometrical
 factors, but not change the underlying linear form of the
 cantilever--dot interaction.   
 
In this appendix we carry out an analysis of the coupling between the
 cantilever
and the electron on the dot, and the normal modes of a nanotube cantilever.  
We verify the
  linear dependence of the cantilever energy on the electric field assumed 
 in the text and derive the form of the coupling constant, $\lambda$.
  By examining the mode spectrum of a nanotube cantilever, we 
 confirm the possibility of obtaining a cantilever with a fundamental flexural
 mode of order $10^8$~Hz, and justify our assumption that for any given cantilever it 
 is sufficient to consider just the lowest mode.

\subsection{Cantilever--Dot Coupling}

We begin by determining the relation between the
maximum electric field at the surface of the dot
and the voltage applied to the conducting cantilever, before analysing the 
details of the effect of the electron on the cantilever energy.   

We will consider a cantilever of length $L\sim1\ \mu$m, positioned so
that its tip lies over the centre of the dot and at a height $d\sim
0.1\ \mu$m.\cite{justify} When an extra electron is added to the dot it
will cause a (classical) vertical displacement of the tip $z\ll d$. We treat the
cantilever as a conducting needle, since it's length will be far greater than 
either of the other two dimensions. An applied voltage induces a line
charge density, $\sigma$, on the cantilever, but because the cantilever is 
necessarily finite in length, the charge density is not entirely uniform and so
for an exact treatment we should write the line charge density as $\sigma(x)$ 
where $x$
runs along the length of the cantilever from the tip ($x=0$). However,
for a cantilever with a large enough aspect ratio, the charge density can be 
approximated as constant with only a small error [the
error is of order $\delta$ with $\delta^{-1}=2\ln(L/r)$, where $r$ is
the radius, for a rod shaped cantilever\cite{jack}]. For the cantilever we are
considering the radius may be as small as $1.5$nm, as we discuss
below, and so the error in assuming a uniform charge density will be
less than 10\%, which is acceptable as our aim is to obtain an order
of magnitude estimate for the  interaction strength.    

The charge induced on the cantilever leads to an electric field at the
surface of the dot, 
\begin{equation}
{\mathbf{E}}=\int_0^{L}\frac{\sigma(-x\hat{\mathbf{i}}+d\hat{\mathbf{j}})
{d}x}{4\pi\epsilon_0(x^2+d^2)^{3/2}},
\end{equation} 
taking $\sigma=cV$ where $c$ is an unimportant  constant and $V$
is the voltage applied to the cantilever. The unit vectors are defined
so that $\hat{\mathbf{i}}$ runs along the cantilever away from its tip and
$\hat{\mathbf{j}}$ points down from the tip towards the dot.
Since $L\gg d$, we find
\begin{equation}
|{\mathbf{E}}|\simeq \frac{2^{1/2}cV}{4\pi\epsilon_0 d}.
\end{equation}

We also need to know how the potential energy of the cantilever varies
for small vertical displacements, $z$, about the equilibrium position,
$d_0$, where $d=d_0-z$. For a section of the cantilever of
length $\Delta x$ centered at $x$, the potential energy due to the
displacement is
\begin{equation}
\Phi(x)=\frac{qcV\Delta x}{4\pi\epsilon_0\sqrt{(d_0-z)^2+x^2}},
\end{equation}
where $q=-e$ is the excess charge on the dot. Since $d_0\gg z$, we obtain 
the $z$--dependent part of the potential energy as
\begin{equation}
\Phi_z(x)=z\frac{qcVd_0\Delta x}{4\pi\epsilon_0[d_0^2+x^2]^{3/2}}.
\end{equation} 
This is of course linear in $z$ as we anticipated in our model. However,
in order to obtain the effective coupling between the electric field
and each of the cantilever modes we need to rewrite the $z$--dependent
part of the potential energy in terms of the normal modes of the
cantilever and so we turn now to the mode spectrum of the cantilever.

\subsection{Cantilever Modes}

A single--walled nanotube cantilever  can be obtained with lengths $\sim
1\ \mu$m, diameters $\sim 3$~nm, and a Young's modulus $\sim 1$~TPa.\cite{nano} 
Recent experiments,\cite{vib,modes2} have shown that the
flexural modes of such nanotubes have frequencies which lie in the MHz--GHz
regime. We can treat a nanotube as a rigid hollow rod and so obtain
the frequencies of the flexural modes,\cite{modes}  
\begin{equation}
\omega_i=\frac{\beta^2_i}{2L^2}\sqrt{ \frac{Y (a^2+b^2)}{\rho}},
\end{equation}   
where $a$ and $b$ are the outer and inner  diameters, $L$ the length,
$Y$ the Young's modulus and $\rho$ the density of the tube. The factors
$\beta_i$ arise from geometrical considerations and are the solutions
of the equation $\cos(\beta_i)\cosh(\beta_i)=-1$. 

The energy of the cantilever can be written in terms of its classical normal
modes as 
\begin{equation}
H=\sum_i\left(\frac{1}{2}m\omega^2_i q^2_i 
+\frac{p^2_i}{2m}\right)I_i,
\end{equation}
where 
\begin{equation}
I_i=\frac{1}{L}\int_0^L\Gamma_i^2{d}x
\end{equation}
and
 \begin{eqnarray}
\Gamma_i(L-x)=&&[\cos(\beta_i)+\cosh(\beta_i)][\sin(\beta_i
x/L)-\sinh(\beta_i x/L)]\nonumber\\
&&-[\sin(\beta_i)+\sinh(\beta_i)][\cos(\beta_i x/L)-\cosh(\beta_i x/L)]
\end{eqnarray}
are the vibrational mode
eigenfunctions.\cite{modes} If we modify the definitions of the
canonical variables slightly to define $q_i'=q_iI_i^{1/2}$ and 
$p_i'=p_iI_i^{1/2}$, then the Hamiltonian takes the conventional form 
\begin{equation}
H=\sum_i\left(\frac{1}{2}m\omega^2_i (q'_i)^2 +\frac{(p'_i)^2}{2m}\right).
\end{equation}
We can add in the additional potential energy due to small amplitude
deflections in the electric field by expanding the displacement $z$ in
terms of the normal modes,
\begin{equation}
V(z)=-\sum_i\frac{ e cV d_0}{4\pi\epsilon_0 I_i^{1/2}}
\left(\int_0^L  \frac{\Gamma_i(x){d}x}{[(x-L)^2+d_0^2]^{3/2}}
\right)q'_i, 
\end{equation}
where the origin of $x$ has been shifted.

We can  quantise the full Hamiltonian for the cantilever in the
usual way, and so we are able to associate a position operator
of the form  
\begin{equation}
\hat{q}'_i=\left(\frac{\hbar}{2m\omega_i}\right)^{1/2}(\hat{a}^{\dagger}_i
+\hat{a}_i)
\end{equation}
with each mode.
Thus, it follows that the interaction between the electron on the dot and
the cantilever can be written in the form proposed [Eq.\ (\ref{coup})]
with a  coupling constant between the cantilever and the
dot which depends on the mode we are considering:
\begin{equation}
\lambda_i= e 
E\xi_i\sqrt{\frac{\hbar}{2m\omega_i}},\label{coupling}
\end{equation}
where $\xi_i$ is a dimensionless constant of order unity defined
by the relation
\begin{equation}
\xi_i=\frac{d_0^2}{(2I_i)^{1/2}}
\int_0^L  \frac{\Gamma_i(x){d}x}{[(x-L)^2+d_0^2]^{3/2}}.
\end{equation}

We can see from Eqs. (\ref{tqd}) and (\ref{ex2}) that the
 effect  of a particular cantilever mode on the electron
interference in the which--path device depends on
$(\lambda_i/\hbar\omega_i)^2$, rather than on $\lambda_i$ alone. Thus, the
ratio of the coupling between the fundamental and the $i$th excited mode goes
as $(\omega_0/\omega_i)^{3}=(\beta_0/\beta_i)^{6}$. Because the electric field
is strongly limited, we will always be working in the regime where the
coupling constant is just sufficient to cause detectable effects. 
Therefore, since $(\beta_0/\beta_1)^6\sim0.005$, our assumption that only
the fundamental mode is relevant is  justified.   

As a concrete example, we consider a nanotube cantilever of length 
$1.4$~$\mu$m and 
outer radius $3.3$~nm.\cite{nano}
In this case, the fundamental frequency is $140$~MHz and the mass is of order 
$8\times 10^{-20}$~kg. If the tip--dot distance is set at $\sim 
0.1$~$\mu$m, then 
$\xi_0\sim 1.3$ and for a maximum electric field of $10^5$~V/m, the corresponding  
maximum coupling constant $\kappa=\lambda/\hbar\omega_0\sim 3$.

\section{Non--Perturbative Calculation of Environmental Coupling}

In Sec. V, we calculated the transmission amplitude through the dot, 
including the effect of the cantilever's environment, by expanding to 
second order in the interaction between the cantilever and 
the electron on the dot, $\lambda$, and then re-exponentiating.
In  this appendix, we justify this result with a full calculation.
The fact that a second order expansion in $\lambda$ leads to the exact 
result is at first somewhat surprising. 
However, the important step is the re-exponentiation of the truncated series 
expansion (\ref{if}): we implicitly equate the 
influence functional with not just the first few terms in an expansion, but 
with an infinite series of terms which are themselves 
composed of products of the second order terms we evaluated. This procedure is 
generally acceptable as an approximation in the limit of small $\lambda$, but
 in this particular  case the series generated in fact coincides with the exact form 
 obtained from a linked cluster expansion.\cite{Mahan}

We begin by adopting the notation
\begin{eqnarray*}
\hat{O}(t)&=&-{\mathrm{e}}^{(-i\omega_0- \gamma)t}\int_0^t dt' 
\hat{F}(t'){\mathrm{e}}^{(i\omega_0+\gamma)t'} \\
\hat{O}^{\dagger}(t)&=&-
{\mathrm{e}}^{(i 
\omega_0-\gamma)t}\int_0^t dt' 
\hat{F}^{\dagger}(t'){\mathrm{e}}^{(-i\omega_0+\gamma)t'},
\end{eqnarray*}
so that 
\begin{eqnarray*}
\hat{a}(t)&=&\hat{a}(0){\mathrm{e}}^{(-i\omega_0-\gamma)t}+\hat{O}(t) \\
\hat{a}^{\dagger}(t)&=&\hat{a}^{\dagger}(0){\mathrm{e}}^{(+i\omega_0-\gamma)t}+
\hat{O}^{\dagger}(t).
\end{eqnarray*}
The operators $\hat{O}(t)$ and $\hat{O}^{\dagger}(t)$ operate only on the 
variables of the oscillator bath and they both commute with $\hat{a}(0)$ and 
$\hat{a}^{\dagger}(0)$ which operate on the states of the cantilever alone. 

The object which we wish to evaluate is the influence functional
\begin{equation}
\left\langle{\mathrm{T}}_t {\mathrm{e}}^{-i \int_0^{t}W_I(t')dt'/\hbar} 
\right\rangle
=\left\langle {\mathrm{T}}_t 
{\mathrm{e}}^{i\lambda 
\int_0^{t}\left[\hat{a}(0){\mathrm{e}}^{(-i\omega_0-\gamma)t'}+\hat{a}
^{\dagger}(0){\mathrm{e}}^{(+i\omega_0-\gamma)t'}\right]dt'/\hbar}\right\rangle
\left\langle {\mathrm{T}}_t 
{\mathrm{e}}^{i\lambda
 \int_0^{t}\left[\hat{O}(t')+\hat{O}^{\dagger}(t')\right]dt'/
 \hbar}\right\rangle, \label{c2}
\end{equation}
where we have exploited the commutation properties of the operators to 
factor the expression into two terms, which we can now write as $C_1\times C_2$.

The method we use to evaluate $C_1$ and $C_2$ is based on 
the application of Wick's theorem, as described in the books by
 Mahan\cite{Mahan}  
and Louisell.\cite{Lou} We can write the first term as
\begin{equation}
C_1=\left\langle {\mathrm{T}}_t 
{\mathrm{e}}^{i\lambda 
\left[\int_0^{t}\hat{a}(0){\mathrm{e}}^{(-i\omega_0-\gamma)t'}+\hat{a}
^{\dagger}(0){\mathrm{e}}^{(+i\omega_0-\gamma)t'}\right]dt'/\hbar}\right\rangle=
\sum_{n=0}^{\infty}i^n U_n(t),
\end{equation}
where 
\begin{eqnarray*}
U_n(t)&=&\frac{(\lambda/\hbar)^n}{n!}\int_0^t dt_1 \ldots\int_0^t
 dt_n 
\left\langle {\mathrm{T}}_t \left[
\hat{a}(0){\mathrm{e}}^{(-i\omega_0-\gamma)t_1}+\hat{a}
^{\dagger}(0){\mathrm{e}}^{(+i\omega_0-\gamma)t_1}\right]\ldots\right. \\
&&\times\left.\left[\hat{a}(0){\mathrm{e}}^{(-i\omega_0-\gamma)t_n}+\hat{a}
^{\dagger}(0){\mathrm{e}}^{(+i\omega_0-\gamma)t_n}\right]\right\rangle. 
\end{eqnarray*}
An obvious simplification arises from the fact that the averages over odd 
numbers of operators will always vanish and so we can replace $n$ by the even 
index $2m$. 

We now apply Wick's theorem which allows us to write the average of products 
of pairs of operators as products of the averages of pairs of operators. Thus 
for $U_{2m}$, we have
\begin{equation}
U_{2m}=\frac{(\lambda/\hbar)^{2m}i^{2m}}{(2m)!}\int_0^t dt_1 \ldots
\int_0^tdt_{2m} \sum_C\left\{ D_0(t_1-t_i)...D_0(t_j-t_{2m}) \right\}, 
\end{equation}
where the summation is over all possible pairing combinations of the $2m$ time 
labels and 
\begin{equation}
iD_0(t_1-t_2)=\left\langle {\mathrm{T}}_t 
\left[\hat{a}(0){\mathrm{e}}^{(-i\omega_0-\gamma)t_1}+\hat{a}
^{\dagger}(0){\mathrm{e}}^{(+i\omega_0-\gamma)t_1}\right] 
\left[\hat{a}(0){\mathrm{e}}^{(-i\omega_0-\gamma)t_2}+\hat{a}
^{\dagger}(0){\mathrm{e}}^{(+i\omega_0-\gamma)t_2}\right]\right\rangle.
\end{equation}
The summation over all possible combinations for each term allows us to 
re-exponentiate so that $
C_1={\mathrm{e}}^{-\phi_0(t)}$, 
where
\begin{equation}
\phi_0(t)=\frac{i}{2}\left(\frac{\lambda}{\hbar}\right)^2\int_0^tdt_1
\int_0^tdt_2 D_0(t_1-t_2).
\end{equation}
It is important to notice that the exact form of the function of time and 
frequency multiplying the operators $\hat{a}(0)$ and $\hat{a}^{\dagger}(0)$ is 
unimportant. These functions give the operators their individual time labels, 
but because they are just algebraic functions they do not affect the validity of 
Wick's theorem.

Now we must consider the second term,
\begin{equation}
C_2=\left\langle{\mathrm{e}}^{i\lambda\int_0^t\left[\hat{O}(t)+\hat{O}^{\dagger}
(t)\right]{d}t'/\hbar}\right\rangle. 
\end{equation}
We can make progress by separating out the underlying operators of the 
oscillators in the bath:
\begin{eqnarray}
\hat{O}(t)&=&-{\mathrm{e}}^{(-i\omega_0-\gamma)t}\int_0^tdt'\sum_{\omega}
\hat{A}(\omega,0)g(\omega){\mathrm{e}}^{-i\omega t'}
{\mathrm{e}}^{(i\omega_0 +\gamma)t} \\
&=&\sum_{\omega}\hat{A}(\omega,0)g(\omega)\left[-{\mathrm{e}}^{(-i\omega_0
-\gamma)t} \int_0^tdt'{\mathrm{e}}^{-i\omega t'}
{\mathrm{e}}^{(i\omega_0 +\gamma)t'}\right] \\
&=&\sum_{\omega}\hat{A}(\omega,0)g(\omega)f(\omega,\omega_0,\gamma,t)
\end{eqnarray} 
and similarly,
\begin{equation}
\hat{O}^{\dagger}(t)=\sum_{\omega}\hat{A}^{\dagger}
(\omega,0)g^*(\omega)f^*(\omega,\omega_0,\gamma,t).
\end{equation}
We do not need to calculate $f$ explicitly since it is just an algebraic
function and so always commutes. Because the oscillators in the heat bath 
are all non--interacting, almost all the $\hat{A}(\omega,0)$ and 
$\hat{A}^{\dagger}(\omega,0)$ operators commute. The only exceptions are 
annihilation and creation operators of the same frequency. This means that we 
can write the expectation value as a product over all the frequencies,
\begin{equation}
C_2=\prod_{\omega}\left\langle 
{\mathrm{e}}^{i\lambda\int_0^t\left[\hat{A}(\omega,0)g(\omega)f+\hat{A}^{\dagger}
(\omega,0)g^*(\omega)f^*\right]dt'/\hbar}\right\rangle. 
\end{equation}

The advantage of decoupling $C_2$ into a product of terms is that each of these 
terms can be handled in the same way as $C_1$. Because $\hat{A}(\omega,0)$ and 
$\hat{A}^{\dagger}(\omega,0)$ are boson operators, Wick's theorem can again be 
applied so that eventually we obtain
\begin{equation}
C_2=\prod_{\omega}{\mathrm{e}}^{-\phi_{\omega}(t)},
\end{equation}
where
\begin{equation}
\phi_{\omega}(t)=\frac{i\lambda^2}{2\hbar^2}\int_0^tdt_1
\int_0^tdt_2 D_{\omega}(t_1-t_2),
\end{equation}
with 
\begin{eqnarray}
iD_{\omega}(t_1-t_2)&=&\left\langle {\mathrm{T}}_t 
\left[\hat{A}(\omega,0)g(\omega)f(\omega,\omega_0,\gamma,t_1)+\hat{A}
^{\dagger}(\omega,0)g^*(\omega)f^*(\omega,\omega_0,\gamma,t_1)\right]\right. \\ \nonumber
&&\times\left. 
\left[\hat{A}(\omega,0)g(\omega)f(\omega,\omega_0,\gamma,t_2)+\hat{A}
^{\dagger}(\omega,0)g^*(\omega)f^*(\omega,\omega_0,\gamma,t_2)\right]\right\rangle.
\end{eqnarray}

The overall expression for the influence functional can now be written in a 
simplified form
\begin{equation}
\left\langle{\mathrm{T}}_t {\mathrm{e}}^{-i 
\int_0^{t}W_I(t')dt'/\hbar}\right\rangle=
{\mathrm{e}}^{-\left[\phi_0(t)+\sum_{\omega}\phi_{\omega}(t)\right]}
={\mathrm{e}}^{-\phi(t)},
\end{equation}
where
\begin{equation}
\phi(t)=\frac{i\lambda^2}{\hbar^2}\int_0^tdt_1
\int_0^{t_1}dt_2 \left\{D_0(t_1-t_2)+\sum_{\omega}D_{\omega}
(t_1-t_2)\right\}.
\end{equation}
However, because the operators in the bath are independent, only the averages 
including  
pairs of operators from the same oscillator are non--zero. Thus, we can simplify 
the sum over $D_{\omega}(t)$ functions: 
\begin{eqnarray}
\sum_{\omega}D_{\omega}(t_1-t_2)&=&\sum_{\omega}\left\langle 
\left[\hat{A}(\omega,0)g(\omega)f(\omega,\omega_0,\gamma,t_1)+\hat{A}^{\dagger}
(\omega,0)g^*(\omega)f^*(\omega,\omega_0,\gamma,t_1)\right]\right.\\ \nonumber
&&\left.\times\left[\hat{A}(\omega,0)g(\omega)
f(\omega,\omega_0,\gamma,t_2)+\hat{A}^{\dagger}
(\omega,0)g^*(\omega)f^*(\omega,\omega_0,\gamma,t_2)\right]
\right\rangle\\
&=&\left\langle  
\left[\hat{O}(t_1)+\hat{O}^{\dagger}(t_1)\right]
\left[\hat{O}(t_2)+\hat{O}^{\dagger}(t_2)\right] 
\right\rangle.
\end{eqnarray}

Since the $\hat{O}(t)$ and $\hat{O}^{\dagger}(t)$ operators commute with 
$\hat{a}(0)$ and $\hat{a}^{\dagger}(0)$ we can complete the process of 
recombination to obtain
\begin{equation}
i\left\{D_0(t_1-t_2)+\sum_{\omega}D_{\omega}(t_1-t_2)\right\}
=\left\langle\left[\hat{a}(t_1)+\hat{a}^{\dagger}(t_1)\right]
\left[\hat{a}(t_2)+\hat{a}^{\dagger}(t_2)\right]\right\rangle. \label{dv}
\end{equation}

We can now use our previous results in Eqs. (\ref{a}), (\ref{b}), and (\ref{c}) 
to evaluate the averages in Eq. (\ref{dv}):
\begin{equation}
\left\langle\left[\hat{a}(t_1)+\hat{a}^{\dagger}(t_1)\right]
\left[\hat{a}(t_2)+\hat{a}^{\dagger}(t_2)\right]\right\rangle
={\mathrm{e}}^{-\gamma (t_1-t_2)}\left[ (\overline{n}+1)
{\mathrm{e}}^{-i\omega_0 (t_1-t_2)}+\overline{n}
{\mathrm{e}}^{i\omega_0 (t_1-t_2)}\right].
\end{equation}
Thus, our final expression for the influence functional  is
\begin{equation}
\left\langle{\mathrm{T}}_t {\mathrm{e}}^{-i \int_0^{t}W_I(t')dt'/\hbar} 
\right\rangle
={\mathrm{e}}^{-\phi(t)},
\end{equation}
with
\begin{equation}
\phi(t)=\frac{\lambda^2}{\hbar^2}\int_0^tdt_1
\int_0^{t_1}dt_2 {{\mathrm{e}}}^{-\gamma (t_1-t_2)}\left[ 
(\overline{n}+1)
{\mathrm{e}}^{-i\omega_0 (t_1-t_2)}+\overline{n}
{\mathrm{e}}^{i\omega_0 (t_1-t_2)}\right]. \label{phic}
\end{equation}
Comparing Eq. (\ref{phic}) with Eq. (\ref{phi}), it is clear that the expression
 we obtained for the influence functional in Sec. V is exact.

\begin{figure}
\caption{Magnitude of the transmission amplitude, $|\langle t_{\rm QD} \rangle|$
at resonance, for a
  cantilever with $\Gamma/\hbar\omega_0=10$ as a function of the dimensionless
  coupling constant
$ \kappa=\lambda/\hbar\omega_0$. The dashed curve is obtained without
an average over incident electron energies, while the full curve includes
the averaging. The amplitudes are normalized to one at $\kappa=0$.  }
\label{fig:one}
\end{figure}

\begin{figure}
\caption{Magnitude of the resonant transmission amplitude, $|\langle 
t_{\mathrm QD}\rangle|$,
 as a function of coupling constant $\kappa$, for  
$\Gamma/\hbar\omega_0$=2 (dotted curve), 1 (dashed curve) and 0.5 (full curve).}
\label{fig:two}
\end{figure}

\begin{figure}
\caption{(a) Magnitude of the transmission amplitude, $|\langle 
t_{\mathrm QD}\rangle |$, and
  (b) the phase, for a
  cantilever with $\Gamma/\hbar\omega_0=0.5$  as a function of coupling constant
$ \kappa$ and the energy detuning 
$\epsilon-\epsilon_{0}$. The average over the electron energy
  distribution has not been performed.}
\label{fig:three}
\end{figure}

\begin{figure}
\caption{(a) Magnitude of the transmission amplitude, $|\langle t_{\mathrm QD}\rangle|$, and
  (b) the phase, for a
  cantilever with $\Gamma/\hbar\omega_0=0.5$  as a function of coupling 
  constant, including
  the effect of the cantilever's environment with 
  $Q=50$, as a function of $\kappa$ and  
  $\epsilon-\epsilon_{0}$.}
\label{fig:four}
\end{figure}

\begin{figure}
\caption{Magnitude of the resonant transmission amplitude, $|\langle 
t_{\mathrm QD}\rangle|$, 
 including the effect of the cantilever's environment as a function of 
 $\kappa$, for  
$\Gamma/\hbar\omega_0=0.5$ and $Q=50$ (dotted curve)
 and  500
(dashed curve). The latter curve has been shifted upwards by 0.5 and 
the former shifted upwards by 1 for clarity.
The case without environmental coupling where  
$\gamma_{c}/\omega_0$=0 (full curve) is included for comparison.}
\label{fig:five}
\end{figure}

\vfill
\eject
\mbox{\epsfig{file=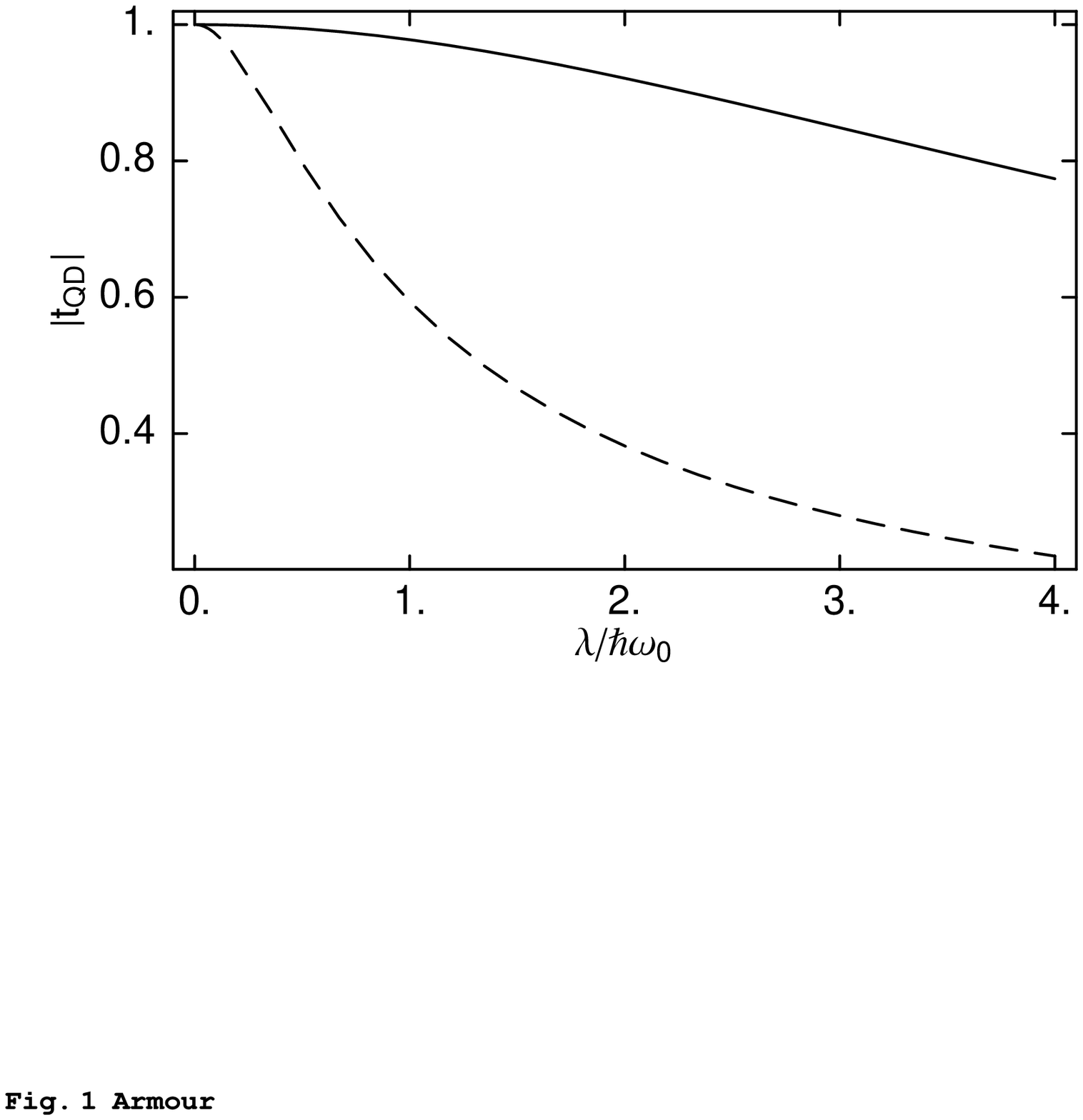, width=6in}}
\mbox{\epsfig{file=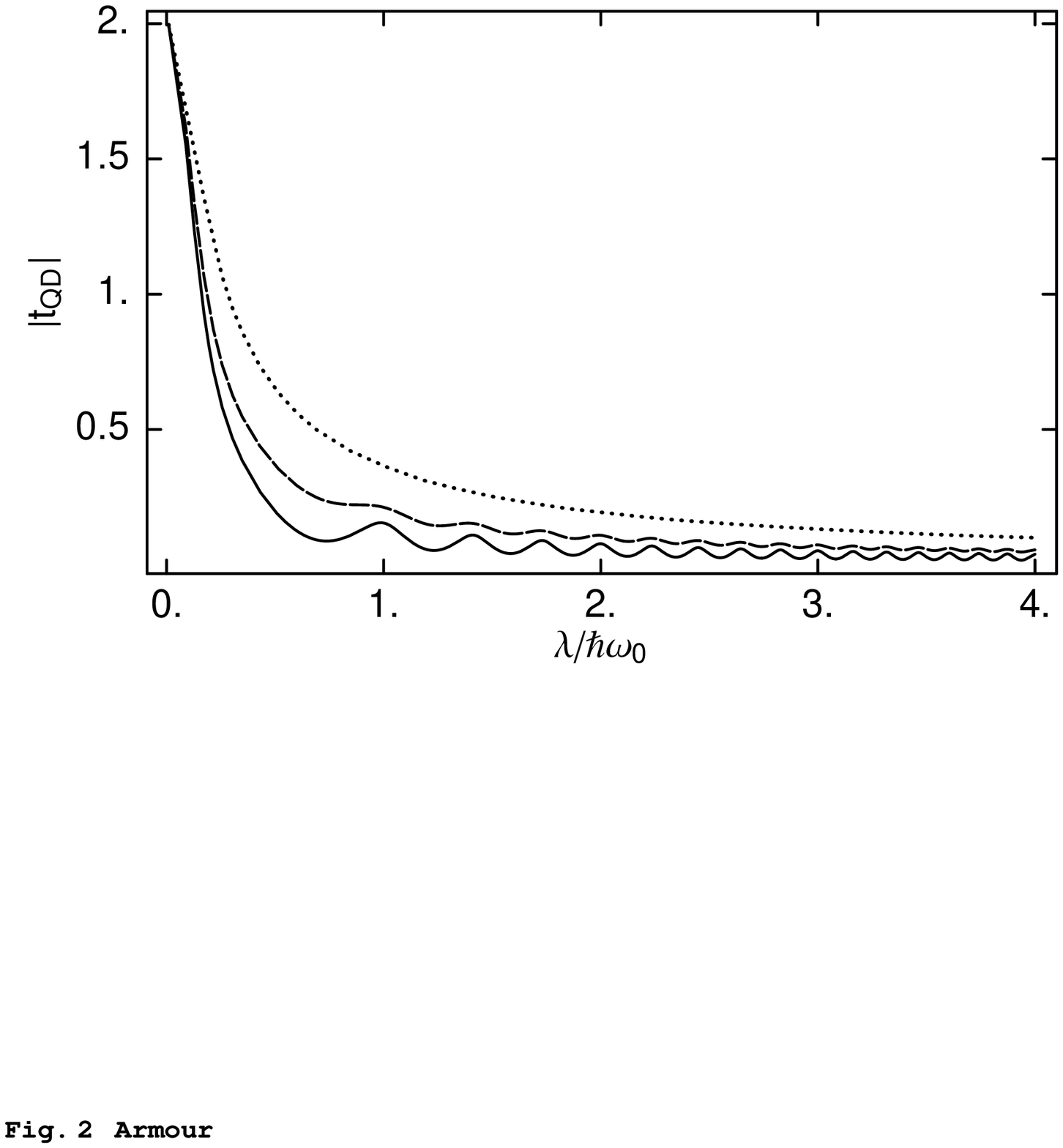, width=6in}}
\mbox{\epsfig{file=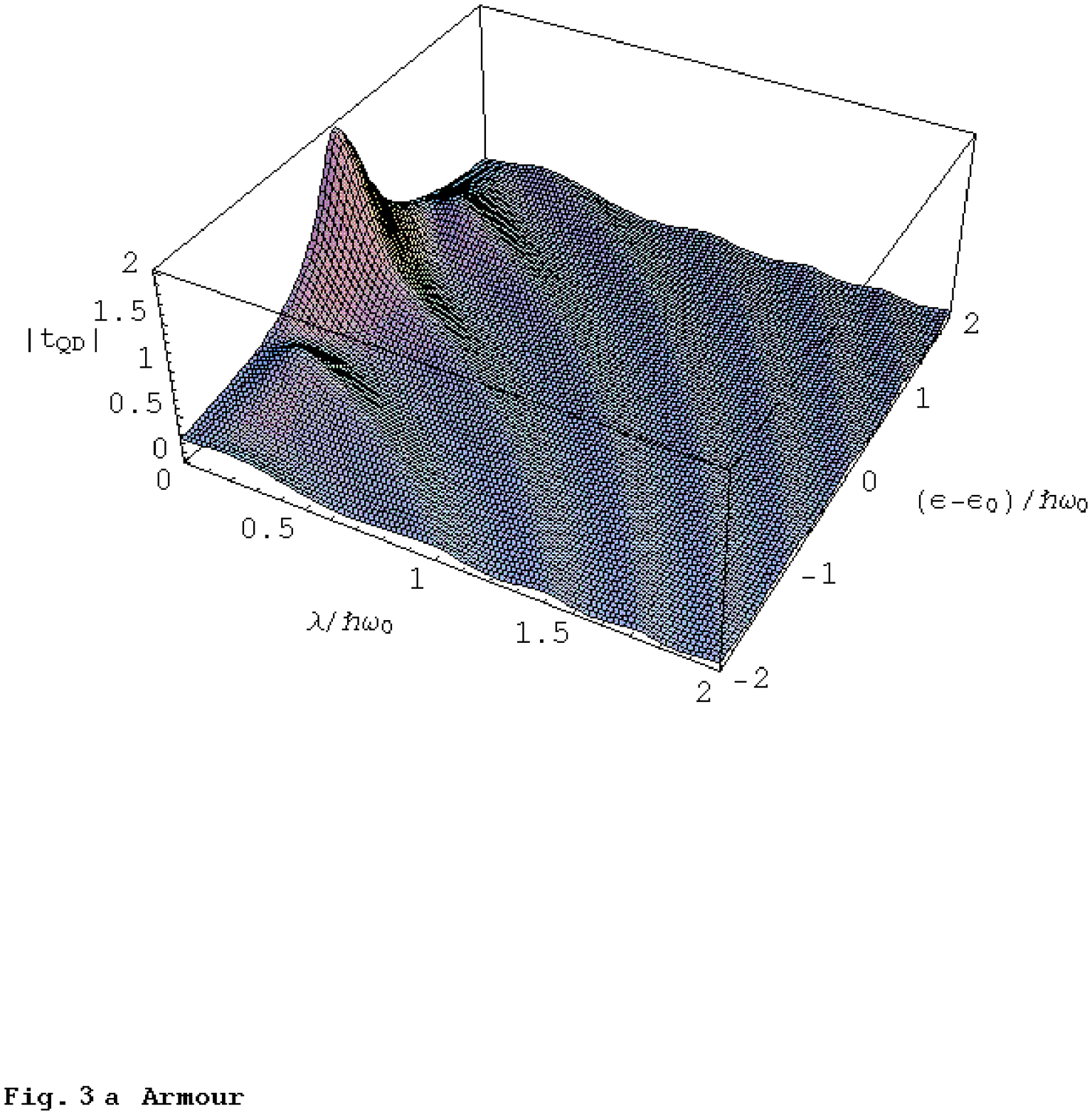, width=6in}}
\mbox{\epsfig{file=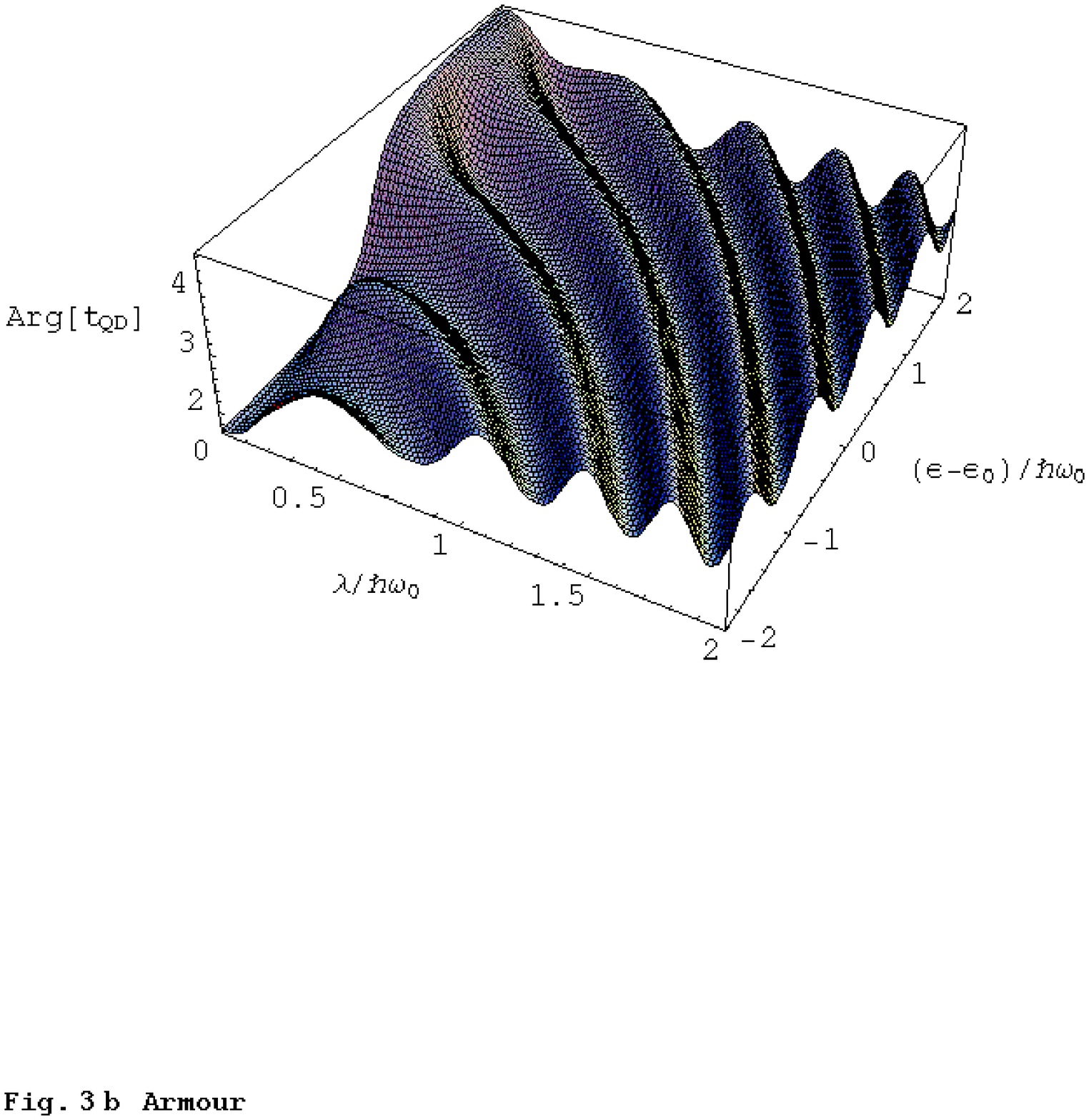, width=6in}}
\mbox{\epsfig{file=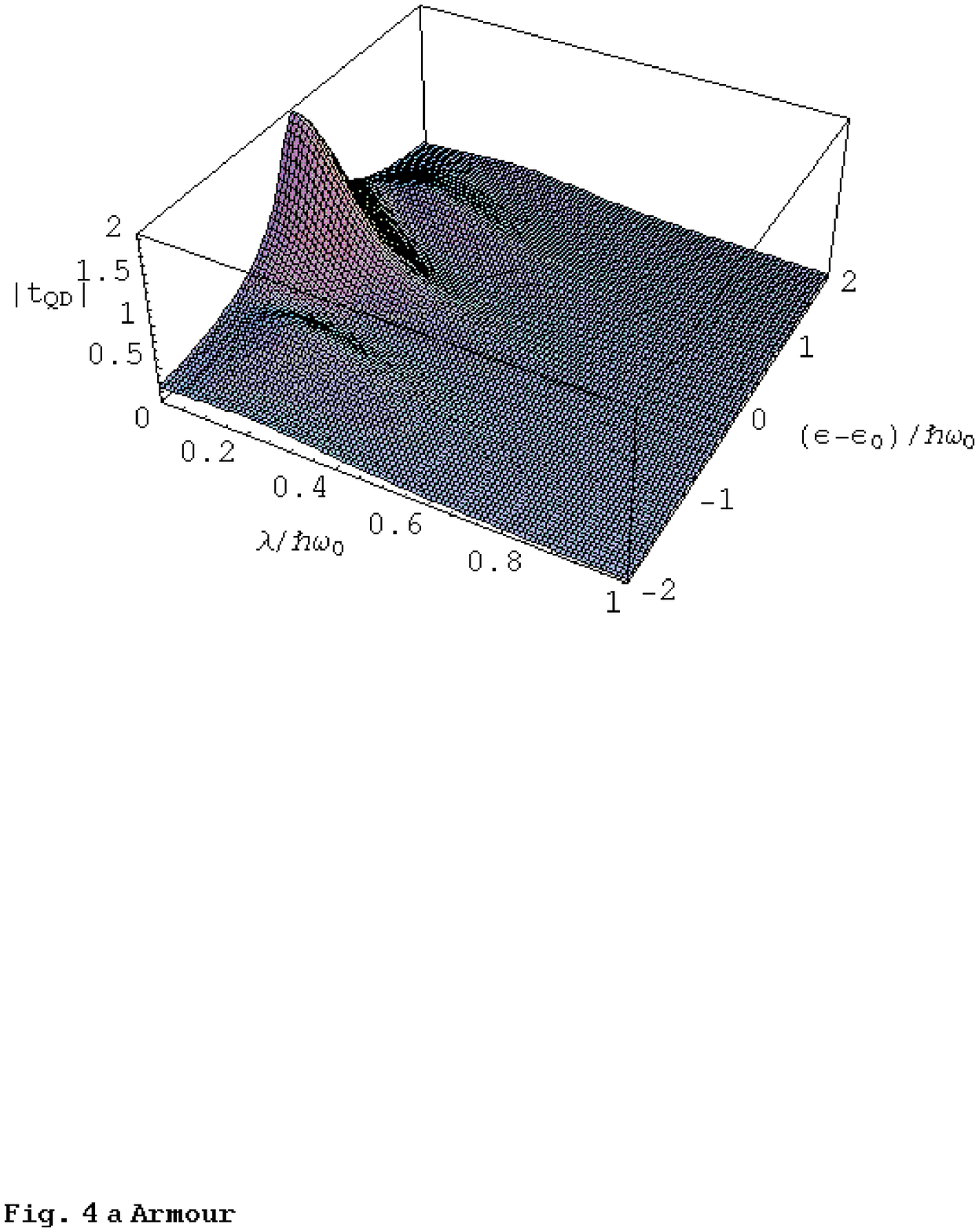, width=6in}}
\mbox{\epsfig{file=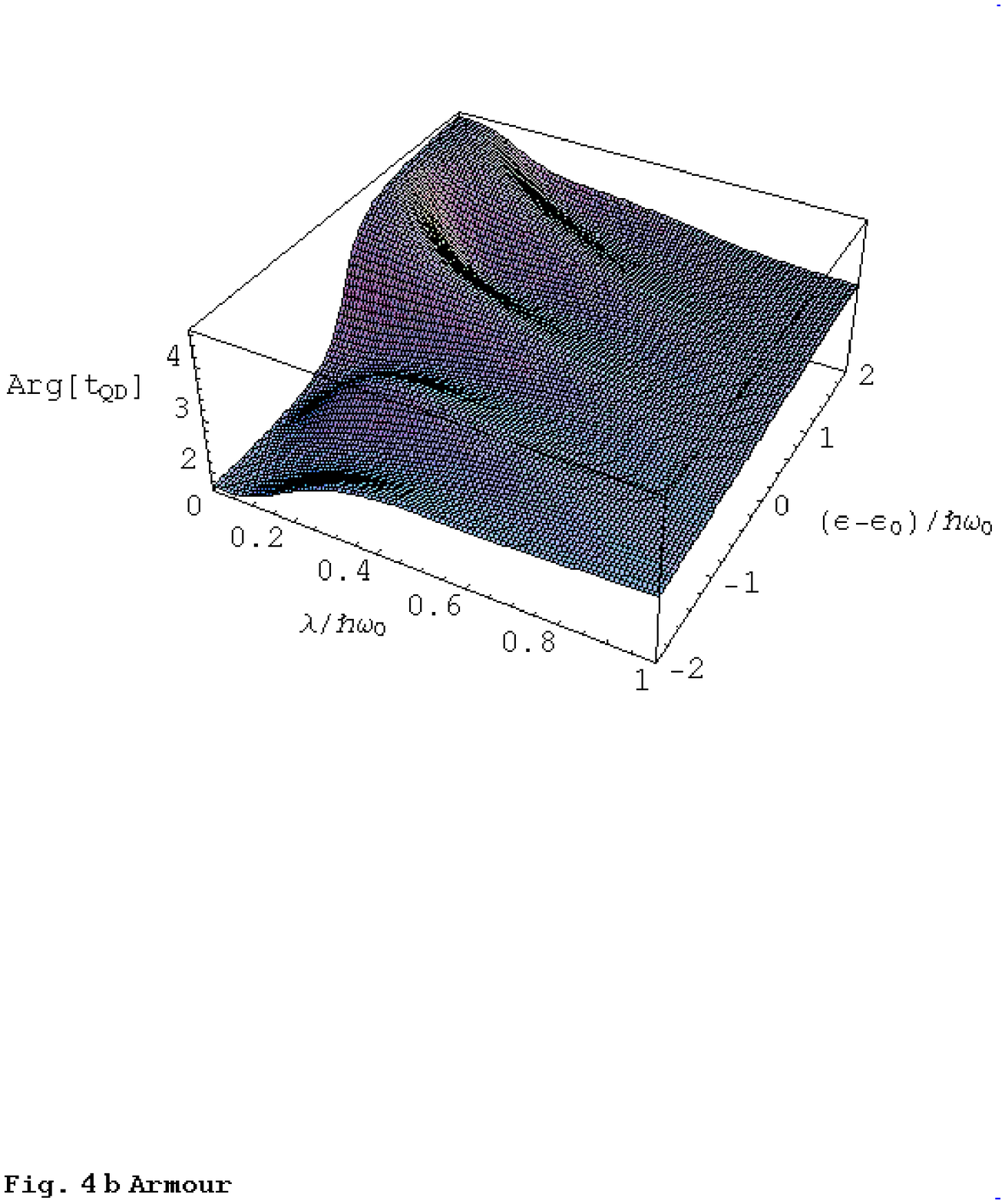, width=6in}}
\mbox{\epsfig{file=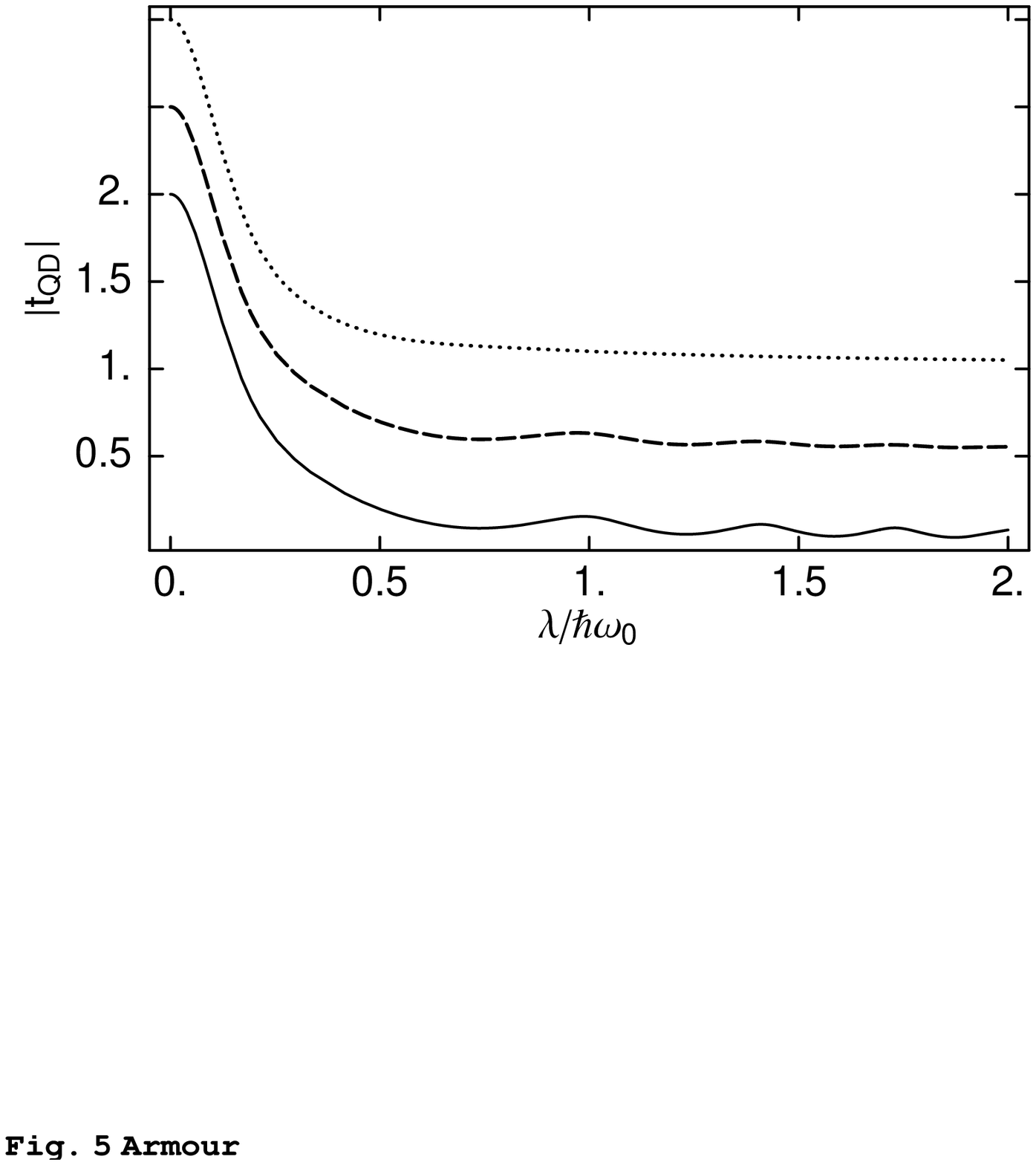, width=6in}}


\begin{references}
\bibitem{steady}Our assumption is that the which--path device is 
operated in a steady--state mode, whereby measurements are only made 
after any transitory behavior has decayed so that the probability 
distribution of (initial) cantilever states is stationary. 
This steady--state distribution 
is further assumed to be the usual thermal distribution. These assumptions will 
only
be strictly correct in the limit as the current goes to zero, but can be expected 
to provide  predictions which are at least qualitatively correct for the average
 behavior over long times  even with 
finite currents.
\bibitem{justify} We assume that all positions are measured from the
  equilibrium height of the cantilever above the dot, including the
  effect of the residual electrons on the dot which do not participate 
 in the transport.
\end{references}
\end{document}